\documentclass[twocolumn]{aastex62}

\usepackage{amsmath}
\usepackage{natbib}
\usepackage{multirow}
\usepackage{color}
\usepackage{bm}
\usepackage[normalem]{ulem}
\usepackage{comment}
\usepackage{tablefootnote}
\usepackage{hologo} 
\usepackage{booktabs} 
\usepackage{threeparttable} 
\usepackage[bf,SL,BF]{subfigure}
\usepackage{afterpage}
\newcommand{\figbox}[1]{%
	\fbox{%
    		\vbox to 1in{%
    		\vfil
    			\hbox to 2in{%
      			\hfil
      			#1%
      			\hfil}%
    		\vfil}}}

\newcommand{\Msun}{M_{\odot}}

\newcommand{\Mej}{M_{\rm ej}}

\def\gsim{\mathrel{\rlap{\lower 4pt \hbox{\hskip 1pt $\sim$}}\raise 1pt \hbox {$>$}}}
\def\lsim{\mathrel{\rlap{\lower 4pt \hbox{\hskip 1pt $\sim$}}\raise 1pt \hbox {$<$}}}

\newcommand{\CenterRow}[2]{
  \dimen0=\ht\strutbox%
  \advance\dimen0\dp\strutbox%
  \multiply\dimen0 by#1%
  \divide\dimen0 by2%
  \advance\dimen0 by-.5\normalbaselineskip%
  \raisebox{-\dimen0}[0pt][0pt]{#2}}
  
\newcommand{\HSC}{HSC-SSP transient survey}

\newcommand{\phz}{photo-$z$}
\newcommand{\spz}{spec-$z$}

\shorttitle{Rapid transients from Subaru HSC-SSP}
\shortauthors{Toshikage et al.}

\begin{document}

\title{A systematic search for rapid transients in the Subaru HSC-SSP transient survey}

\correspondingauthor{Seiji Toshikage}
\email{seiji.toshikage@astr.tohoku.ac.jp}

\author[0009-0002-5156-7819]{Seiji Toshikage}
\affiliation{Astronomical Institute, Tohoku University, Sendai 980-8578, Japan}

\author[0000-0001-8253-6850]{Masaomi Tanaka}
\affiliation{Astronomical Institute, Tohoku University, Sendai 980-8578, Japan}
\affiliation{Division for the Establishment of Frontier Sciences, Organization for Advanced Studies, Tohoku University, Sendai 980-8577, Japan}

\author{Naoki Yasuda}
\affiliation{Kavli Institute for the Physics and Mathematics of the Universe (WPI), The University of Tokyo Institutes for Advanced Study, The University of Tokyo, 5-1-5 Kashiwanoha, Kashiwa, Chiba 277-8583, Japan}

\author[0000-0003-1169-1954]{Takashi J. Moriya}
\affiliation{National Astronomical Observatory of Japan, National Institutes of Natural Sciences, 2-21-1 Osawa, Mitaka, Tokyo 181-8588, Japan}

\author[0000-0003-2691-4444]{Ichiro Takahashi}
\affiliation{Department of Physics, Tokyo Institute of Technology, Meguro-ku, Tokyo 152-8551, Japan}

\author[0000-0002-9092-0593]{Ji-an Jiang}
\affiliation{Department of Astronomy, University of Science and Technology of China, Hefei 230026, China}
\affiliation{National Astronomical Observatory of Japan, National Institutes of Natural Sciences, 2-21-1 Osawa, Mitaka, Tokyo 181-8588, Japan}

\author[0000-0001-6402-1415]{Mitsuru Kokubo}
\affiliation{National Astronomical Observatory of Japan, National Institutes of Natural Sciences, 2-21-1 Osawa, Mitaka, Tokyo 181-8588, Japan}

\author[0000-0002-8299-0006]{Naoki Matsumoto}
\affiliation{Astronomical Institute, Tohoku University, Sendai 980-8578, Japan}

\author[0000-0003-2611-7269]{Keiichi Maeda}
\affiliation{Department of Astronomy, Kyoto University, Kitashirakawa-Oiwake-cho, Sakyo-ku, Kyoto 606-8502, Japan}

\author[0000-0001-7449-4814]{Tomoki Morokuma}
\affiliation{Astronomy Research Center, Chiba Institute of Technology, 2-17-1 Tsudanuma, Narashino, Chiba 275-0016, Japan}
\affiliation{Planetary Exploration Research Center, Chiba Institute of Technology, 2-17-1 Tsudanuma, Narashino, Chiba 275-0016, Japan}

\author[0000-0001-7266-930X]{Nao Suzuki}
\affiliation{E.O. Lawrence Berkeley National Laboratory, 1 Cyclotron Rd., Berkeley, CA, 94720, USA}
\affiliation{Department of Physics, Florida State University, 77 Chieftan Way, Tallahassee, FL 32306, USA}
\affiliation{Laboratoire de Physique Nucleaire et de Hautes-Energies, 4 Place Jussieu, 75005 Paris, France}

\author[0000-0001-8537-3153]{Nozomu Tominaga}
\affiliation{National Astronomical Observatory of Japan, National Institutes of Natural Sciences, 2-21-1 Osawa, Mitaka, Tokyo 181-8588, Japan}




\begin{abstract}
Recent high-cadence transient surveys have discovered rapid transients whose light curve timescales are shorter than those of typical supernovae. In this paper, we present a systematic search for rapid transients at  medium-high redshifts among 3381 supernova candidates obtained from the Subaru \HSC. We developed a machine learning~classifier to classify the supernova candidates into four types~(Type Ia, Ibc, II supernovae, and rapid transients) based on the features derived from the light curves. By applying this classifier to the 3381 supernova candidates and by further applying the quality cut, we selected 14 rapid transient samples. They are located at a wide range of redshifts~($0.34 \leq z \leq 1.85$) and show a wide range of the peak absolute magnitude~($-17 \geq M \geq -22$). The event rate of the rapid transients is estimated to be $\sim 6\times10^3 ~\rm{events~yr^{-1}~Gpc^{-3}}$ at $z \sim 0.74$, which corresponds to about $2$~\% of the event rate of normal core-collapse supernovae at the similar redshift. Based on the luminosity and color evolution, we selected two candidates of Type Ibn supernovae at $z\sim0.75$. The event rate of Type Ibn SN candidates is more than 1 \% of Type Ib SN rate at the same redshift, suggesting that this fraction of massive stars at this redshift range eruptively ejects their He-rich envelope just before the explosions.
Also, two objects at $z=1.37$ and 1.85 show high luminosities comparable to superluminous supernovae. Their event rate is about 10-25 \% of superluminous supernovae at $z\sim 2$.
\end{abstract}



\section{Introduction}
\label{sec:intro}

In the last decade, high-cadence and deep optical transient surveys such as the Palomar Transient Factory~(PTF;~\citealt{law2009, rau2009}), the Pan-STARRS1~(PS1;~\citealt{kaiser2010}), All Sky Automated Survey for SuperNovae~(ASAS-SN;~\citealt{shappee2014}),  Asteroid Terrestrial-impact Last Alert System~(ATLAS;~\citealt{tonry2018}), and the Zwicky Transient Facility~(ZTF;~\citealt{bellem2019}) have dramatically increased the number of discovered supernovae~(SNe).
These surveys have also revealed a great diversity of SNe in luminosity and timescale. For example, superluminous SNe~(SLSNe;~\citealt{quimby2011, gal-yam2012, moriya2018}) is one of the rare populations of SNe discovered by those surveys which are characterized by the peak absolute magnitude brighter than $\sim -21$ mag. 

From a perspective of timescale, recent surveys have discovered new types of transients whose timescale is shorter than those of typical SNe. These transients are often called ``rapid transient'' or ``rapidly evolving transient''. 
Rapid transients have been explored by various surveys such as PS1 (\citealt{drout2014}), PTF \citep{whitesides2017}, the Dark Energy Survey~(DES;~\citealt{pursiainen2018}), the Supernova Legacy Survey~(SNLS;~\citealt{arcavi2016}), the Subaru HSC survey ~\citep{tanaka2016, tominaga2019, tampo2020}, and the ZTF high cadence survey~\citep{ho2023}. 
From these observations, it has been revealed that rapid transients have a large variety in their luminosities and timescales. 

Rapid transient can be mainly dominated by three types of transients~\citep{ho2023} namely (1)~Type IIb SNe, (2)~Type Ibn SNe, and (3)~AT2018cow-like transients. Low luminous~($M_{\rm{peak}}>-18$) rapid transients are dominated by Type IIb SNe while intermediately luminous~($-20>M_{\rm{peak}}>-18$) rapid transients are dominated by Type Ibn SNe. 
In the case of Type IIb SNe, the bright first peak by the shock cooling emission can show the rapidly evolving light curves~\citep{fremling2019}. On the other hand, in the case of Type Ibn SNe, which are characterized by the strong and narrow He emission lines in their early spectrum \citep{pastorello2015, hosseinzadeh2018}, interaction with circumstellar material (CSM) can produce the light curves with luminous peaks and rapid evolutions~\citep{maeda2022, takei2024}. In a high luminosity~($M_{\rm{peak}}<-20$) and short timescale regime, so called luminous fast blue optical transients~(LFBOTs) have been discovered, such as KSN 2015K~\citep{rest2018}, AT2018cow~\citep{prentice2018, perley2019}, CSS161010~\citep{coppejians2020}, AT2018lug~\citep{ho2020}, AT2020xnd~\citep{perley2020}, MUSSES2020J~\citep{jiang2022}, AT2023fhn~\citep{chrimes2024}. 
Although various scenarios have been proposed for the origin of LFBOTs such as CSM interaction~\citep{fox2019, pellegrino2022}, central engine including spin down energy of magnetar or accretion to the black hole~\citep{prentice2018, mohan2020}, tidal disruption events~\citep{perley2019, kuin2019, metzger2022}, the nature of LFBOTs remains unclear. 

The event rate of each subtype of rapid transients is of interest to understand the progenitor stars of these transients. 
Although the event rates of rapid transients were estimated in previous works~\citep{drout2014, pursiainen2018, tampo2020}, these event rates included not only the rare transients such as Type Ibn SNe and LFBOTs but also Type IIb SNe which originate from normal core-collapse SNe. \cite{ho2023} estimated the event rate for each type of rapid transients separately for the first time.
The event rate of LFBOTs was obtained as $70^{+350}_{-68}~\rm{events~yr^{-1}~Gpc^{-3}}$, which is about 0.1\% of core-collapse SNe. 
However, the redshift evolution of the event rates remains unknown because most of the samples in \cite{ho2023} are located at low redshift~($z \leq 0.3$).

In this work, we explore rapid transient from the 3381 supernova candidates obtained by the Hyper~Suprime-Cam~Subaru~Strategic~Program~transient~survey~(the Subaru HSC-SSP transient survey;~\citealt{aihara2018, yasuda2019}). From the rapid transient samples, we select Type Ibn SN candidates with photometric properties such as color evolution and peak luminosity. We discuss the physical properties of our samples and estimate the event rate. This paper is structured as follows. In Section~\ref{sec:hsc}, we give a brief overview of the Subaru HSC-SSP transient survey. In Section~\ref{sec:method}, we present the methods for the classification of rapid transients. In Section~\ref{sec:results}, we show properties of our rapid transient samples. In Section~\ref{sec:discussion}, we discuss properties of the host galaxies and event rates of the rapid transients. Finally we give conclusions in Section \ref{sec:conclusions}. We use the AB magnitude system~\citep{oke1983} throughout this paper.

\section{Summary of the Subaru HSC-SSP transient survey}
\label{sec:hsc}

\begin{table*}[ht!]
    \begin{center}
    \caption{Summary of the Subaru \HSC}
    \label{tab:hscssp}
        \begin{tabular}{ccccccccc}
        \midrule 
        \CenterRow{2}{Field} & \CenterRow{2}{Period} & \CenterRow{2}{Layer} & Area &  \CenterRow{2}{Number of epochs} & \multicolumn{4}{c}{$5\sigma$ limiting magnitudes}\\
         &  &  & (deg$^{2}$)  & & HSC-G & HSC-R2 & HSC-I2 & HSC-Z \\         
        \midrule 
        \CenterRow{2}{COSMOS} & \CenterRow{2}{Nov.~2016~-~Apr.~2017} & Ultra-Deep & 1.8  &  8-12 & 26.4 & 26.3 &  26.0& 25.6\\
        &  & Deep  & 5.8 & 4-8 & 25.8 & 25.8 & 25.5 & 25.0 \\
        \hline
        \CenterRow{2}{SXDS} & \CenterRow{2}{Nov.~2019~-~Mar.~2020} &  Ultra-Deep & 1.8  &  6-10 & 26.3 & 26.0 & 25.6 & 25.2 \\
        &  & Deep & 5.3 & 3-6 & 25.9 & 25.5 & 24.9 & 24.5 \\	
        \hline \\
        \end{tabular}
    \end{center}
\end{table*}

\begin{figure}[t]
    \begin{center}
        \includegraphics[scale=0.6]{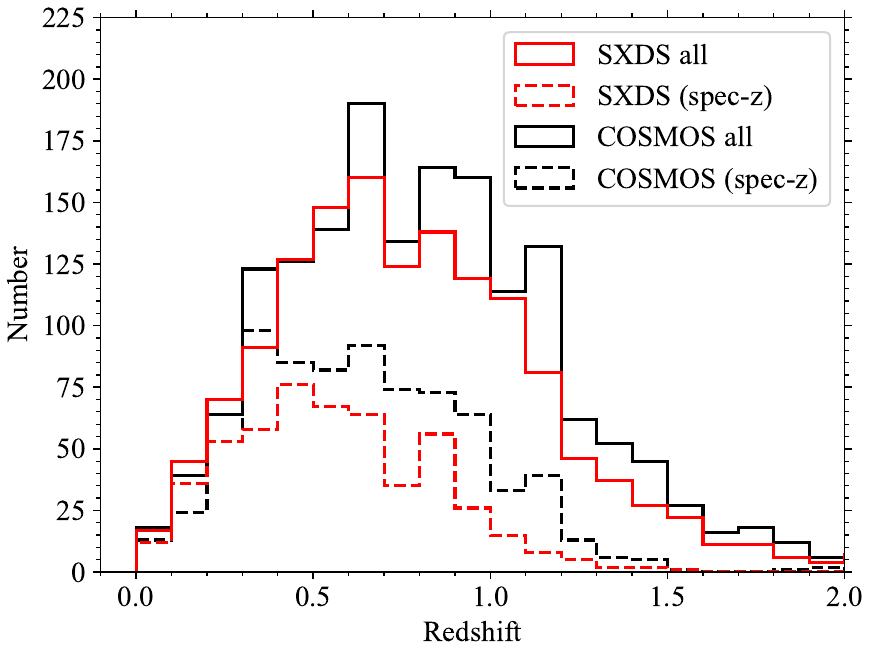}
        	\caption{Distribution of redshifts for our SN samples in the SXDS field~(red lines) and the COSMOS field~(black lines). Dashed lines represent the distribution of SN samples with spectroscopically measured redshifts in each field.
	\label{fig:sxds-zdist}
    	}
    \end{center}
\end{figure}

We analyze the transient samples from the Subaru HSC-SSP survey~\citep{aihara2018}. As a part of the HSC-SSP survey, repeated time-domain observations were performed for the COSMOS (Cosmic Evolution Survey,~\citealt{scoville2007}) and the SXDS field (Subaru/XMM-Newton Deep Survey,~\citealt{furusawa2008}).
For more details of the transient survey of the COSMOS field, see \citet{yasuda2019}.
In this work, we use the data for the COSMOS and the SXDS field.
The data were obtained in \it{}g,~r,~i,~z \rm{and} \it{}y\rm{}~bands \rm{}with the cadence of 7-10 days. The magnitude limits in each field are listed in Table~\ref{tab:hscssp}. Thanks to the deep depth~(25-26~mag) and the large area for this depth~($14.7~\rm{deg^{2}}$ in total), the Subaru HSC-SSP transient survey achieves a large survey volume, which enables us to study rapid transients at a higher redshift range ($z > 0.5$) as compared with the other surveys. Considering the cadence and the depth, our survey is sensitive to the objects with a duration of $\geq 3.5$-$5.0$~$\rm{days}$ at $z=1$.
~Note that although the observations were performed in these five bands, we only use the four bands~($g$, $r$, $i$, $z$) for this work because the $y$-band has a relatively shallow limiting magnitude. The typical value of galactic extinction in the survey fields is $E(B-V) = 0.02$~\citep{schlafly2011}. In this work, we do not correct the effect of Galactic extinction in our analysis because this level of extinction does not largely affect the results.\par
The observational data were processed with the HSC pipeline~\citep{juri2017, ivezi2019, bosch2018}, including bias, dark, flat, and fringe corrections.
Astrometric and photometric calibrations were done against the PS1 catalog \citep{chambers2016, schlafly2012, tonry2012, magnier2013}.
For the reduced images, image subtraction was performed with the deep reference images constructed from the data taken between March 2014 and April 2016 for the COSMOS field and between September 2014 and January 2018 for the SXDS field. After image subtraction, we coadded the difference images for each filter at each epoch. Based on these coadded images, we identified transient sources. We applied a convolutional neural network~(CNN) to classify the detected sources as real or bogus detection. As a result, 65,387 and 45,389 transient candidates were identified in the COSMOS and SXDS fields respectively. 

We identified the putative host galaxy for each candidate as in \citet{yasuda2019}. Basically the closest object in each reference image was labeled as a host galaxy. 
When the matched object is a very faint, noise-like source or a faint source deblended from a big galaxy, the identification was corrected by visual inspections (see also Section \ref{sec:results}). Then, the host galaxies were matched with public redshift catalogs~(COSMOS2015: \citealt{laigle2016}, HSC photo-$z$ catalog: \citealt{tanakamasayuki2018}, and SPLASH photo-$z$ catalog: \citealt{mehta2018}). Among the transient candidates, we excluded the transients with light curves dominated by negative PSF fluxes. Also, we excluded the sources that spatially coincide with point sources, which are likely to be variable stars. Remaining 26,988 and 16,549 transient candidates in COSMOS and SXDS fields were visually inspected. In this classification process, we mainly classified transients into SN, AGN, and SN/AGN (marginal class) based on the offset from the putative host galaxies and light curve shapes. For example, objects with multiple peaks or a very long duration are classified as AGN. AGN and SN/AGN were not included in the supernovae candidates in this work.
We finally obtained 3381~(1824/1557) SN candidates in COSMOS/SXDS fields (see \citealt{yasuda2019} for more  details for the COSMOS field). 
Among the 3381 SN candidates, the redshifts for the host galaxies of 3174 SNe have been obtained. 
Among these objects, the redshifts of 1277 objects have been  spectroscopically measured while the redshifts of the other objects have been derived from photometric redshifts: 255 objects from COSMOS2015 catalog~\citep{laigle2016}, 1628 objects from HSC photometric redshift catalog~\citep{tanakamasayuki2018}, and 14 objects from SPLASH photo-$z$ catalog~\citep{mehta2018}. In Figure \ref{fig:sxds-zdist}, we show the redshift distribution of SNe in each field. For the remaining 207 objects without redshifts, 128 objects are hostless SNe while the photometric redshifts of 79 objects are not be reliably detemined by SED fitting.

\section{selection of rapid transients}
\label{sec:method}
Rapid transients among the Subaru HSC-SSP transient survey in the COSMOS field have been explored by \cite{tampo2020}. In \cite{tampo2020}, observational light curves in each band were fitted by a simple, symmetric Gaussian function.
By using the FWHMs of the light curves as criteria,
5 rapid transients~(HSC17auls, HSC17bbaz, HSC17bhyl, HSC17btum, HSC17dadp) were selected. However, this selection is not necessarily accurate because light curves of SN are typically asymmetric. 
Furthermore, the threshold for the timescale between rapid transients and typical SNe is not distinct due to the continuous distribution of some types of SNe~(e.g., Type IIb and Ibn SNe;~\citealt{ho2023}) 

\begin{figure}[t]
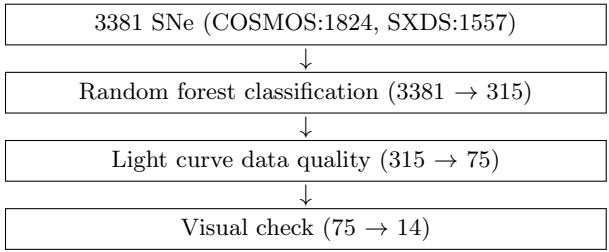

\begin{center}
	\framebox[8cm][c]{3381 SNe (COSMOS:1824, SXDS:1557)}\\
	$\downarrow $ \\
	\framebox[8cm][c]{Random forest classification (3381 $\rightarrow$ 315)}\\
	$\downarrow $ \\
	\framebox[8cm][c]{Light curve data quality (315 $\rightarrow$ 75)}\\
	$\downarrow $ \\
	\framebox[8cm][c]{Visual check (75 $\rightarrow$ 14)}\\
\caption{
\label{fig:select}
A flow chart of the rapid transient selection.}
\end{center}
\end{figure}

Thus, in this work, we overcome these difficulties with a machine learning (ML) based classification technique. We search for rapid transient candidates by the following steps~(Figure~\ref{fig:select}). We first develop a supervised ML model with random forest~(RF) and classify 3381 objects into four types~(Type Ia, Ibc, II SNe, and rapid transients) with features derived from the light curves. 
Then, we further impose various criteria of the data quality. Finally, we select rapid transients by comparing the light curves with those of known rapid transients. 

\subsection{Classification with RF classifier}

In this subsection, we describe the photometric classification methods for our samples with ML classifier. We adopted RF~\citep{breiman2001}, which is a supervised ML method based on the combination of decision trees. For the classification of our samples with RF, we first trained our RF model with training data set consisting of the features derived from mock light curves of Type Ia, Ibc, II SNe, and rapid transients. We generated mock light curves with \texttt{SNCosmo}, a Python Library for Supernova Cosmology~\citep{sncosmo}. 

We used the SALT2~\citep{guy2007, guy2010} model template for Type Ia SNe in which the light curve is expressed with the stretch parameter $x_{1}$ and color parameter $c$. The peak absolute magnitude and these parameters are related by the equation:
\begin{equation}
M_{B} = m_{B} - \mu - \alpha x_{1}  - \beta c.
\end{equation}
We used $\alpha = 0.141$, $\beta = -3.101$, and $m_{B}-\mu = -19.05$ ,where $\mu$ is a distance modulus. The stretch parameter $x_{1}$ was distributed in a range of $-3.0<x_{1}<3.0$ with the mean value of 0.945 and a standard deviation of 1.553/0.257 for the lower/upper sides. The color parameter $c$ was also distributed in a range of $-0.25<c<0.25$ with the mean value of $-0.043$ and a standard deviation of 0.052/0.107 for lower/upper sides. For Type Ibc and Type II SNe, we used the \texttt{SNCosmo} built-in sources~\citep{kessler2009} as observational templates. Based on those templates, we distributed the peak absolute magnitude around $-16.8$/$-17.5$ with the standard deviation of 1.0/1.2 for each Type II/Ibc SNe~\citep{richardson2014}.

\begin{figure}[t]
  \begin{center}
    \includegraphics[scale=0.45]{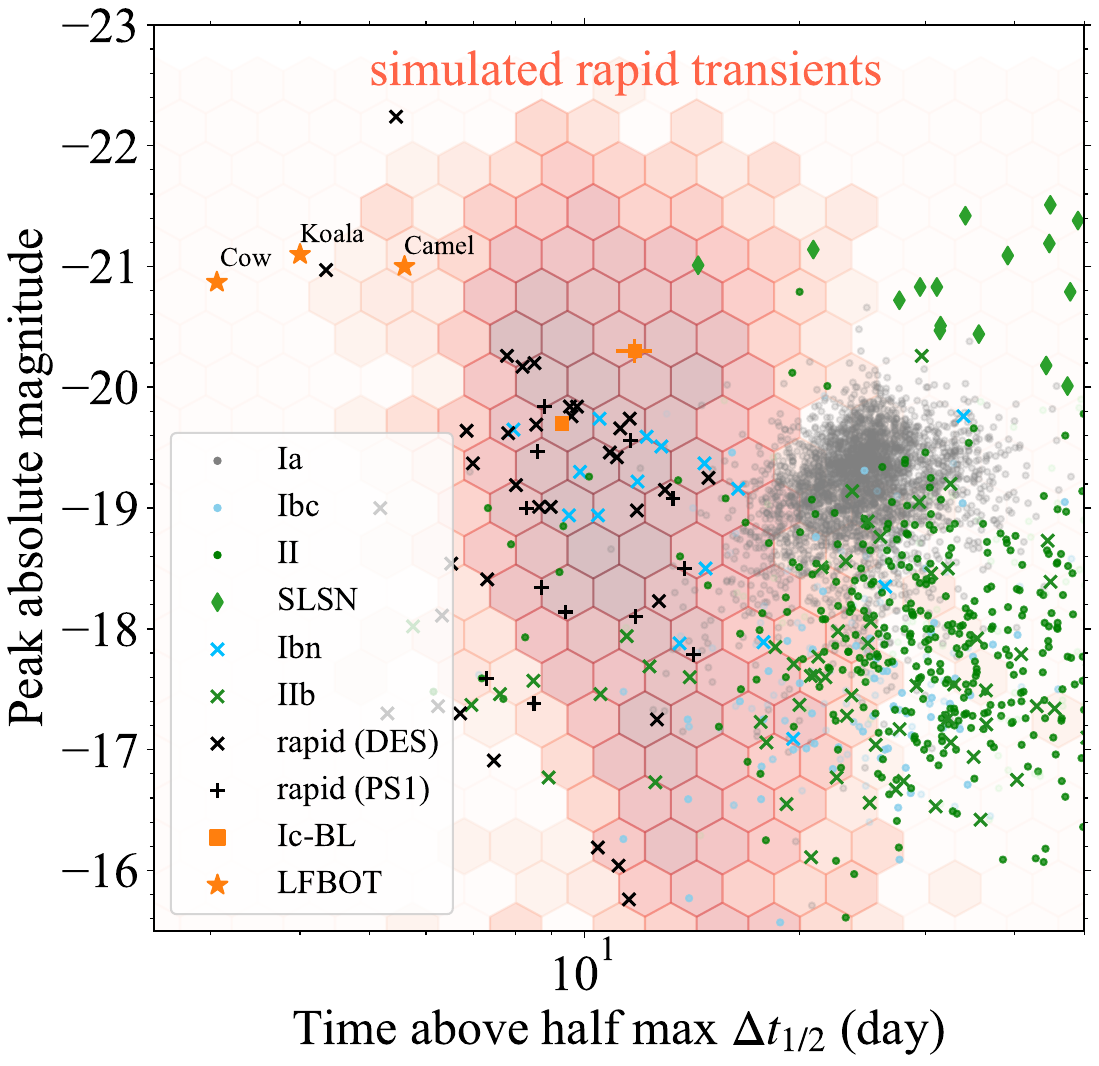}
\caption{Phase diagram of optical transients~\citep{perley2020, pursiainen2018, drout2014} with our simulated rapid transient samples shown as orange shaded region (density plot of the color represents relative numbers of simulated samples).
  \label{fig:phase-sim}
  }
\end{center}
\end{figure}

For rapid transients, the number of the observational samples with a good wavelength coverage is not sufficient to construct model templates. Also, even among the rapid transients, there is a large diversity in their light curves as shown in \cite{ho2023}. Therefore, we created diverse templates through a simple analytical $\rm{^{56}Ni}$-powered model based on \cite{arnett1982}.
Multi-color light curves were calculated 
by simply assuming the blackbody spectral energy distribution. 
It is noted that rapid transients are not necessarily powered by $^{56}$Ni, i.e., some objects require other power sources such as CSM interaction or magnetar/BH accretion. 
However, we generated the various light curves of rapid transients within the framework of the Arnett model by adopting a wide range of physical parameters~($0.1~\Msun \leq \Mej \leq 0.3~\Msun$, $0.1~\Msun \leq M_{\rm{Ni}} \leq 0.3~\Msun$, and  $10,000~\rm{km~s^{-1}}\leq \it{v_{\rm{ej}}} \leq \rm{20,000}~\rm{km~s^{-1}}$).
In fact, we even allowed unrealistic parameter sets (such as $M_{\rm Ni} > M_{\rm ej}$) to produce a wide variety.
In addition, since UV flux of rapid transients do not necessarily follow the blackbody function, we randomly shifted the calculated light curve in magnitude to provide a further diversity in the light curve color.
As a result, properties of this template set is wide enough to cover the properties of all the sub types of rapid transients, i.e.,
the peak absolute magnitude of $-15\geq M_{\rm peak}\geq-22$ and timescale of $5~\rm{days}~$$\leq \Delta t_{1/2}\leq 20$ days (Figure \ref{fig:phase-sim}).

In fact, this wide parameter space overlaps with other types of SNe, and thus, our classifier provides a conservative selection not to miss genuine rapid transients (at the sacrifice of the purity). 

With these model templates, mock observations were performed according to the observational schedule and depth of each field and layer~(COSMOS UD/D, SXDS UD/D) using \texttt{SNCosmo}~\citep{sncosmo}. We generated the 10,000 simulated data for Type Ia, Ibc, and II SNe, and 3,000 simulated data for rapid transients. We used a smaller number for rapid transients by considering their smaller fraction in actual observations.

\begin{table*}
  \caption{Derived features for classification}
  \label{tab:feature}
  \centering
  \begin{tabular}{lll}
  \hline 
label & definition \\
  \hline 
  \hline 
flux\_[min, max]\_[$g$, $r$, $i$, $z$] & minimum and maximum flux in each band [$g$, $r$, $i$, $z$] \\
peak\_dt\_[$g$, $r$, $z$] & difference of peak time in days between [$g$, $r$, $z$] band and $i$ band in rest frame\\
mag\_max\_[$g$, $r$, $i$, $z$]& peak observed magnitude [$g$, $r$, $i$, $z$] \\
lumi\_[$g$, $r$, $i$, $z$]& absolute flux corrected for the distance [$g$, $r$, $i$, $z$] \\
positive\_ratio\_[$g$, $r$, $i$, $z$]& the ratio of the peak flux and the difference between peak and minimum flux in each band [$g$, $r$, $i$, $z$]\\
decline\_time\_[7,5,2]\_[$g$, $r$, $i$, $z$] & decline time from peak to given fraction [0.7, 0.5, 0.2] of flux in each band [$g$, $r$, $i$, $z$] in rest frame\\
rise\_time\_[7,5,2]\_[$g$, $r$, $i$, $z$] & rise time from peak to given fraction [0.7, 0.5, 0.2] of flux in each band [$g$, $r$, $i$, $z$] in rest frame\\
redshift & redshift of the host galaxy~(spec-$z$ or photoz)\\
  \hline
  \end{tabular}
\end{table*}

\begin{figure*}[t]
  \begin{center}
    \includegraphics[scale=0.6]{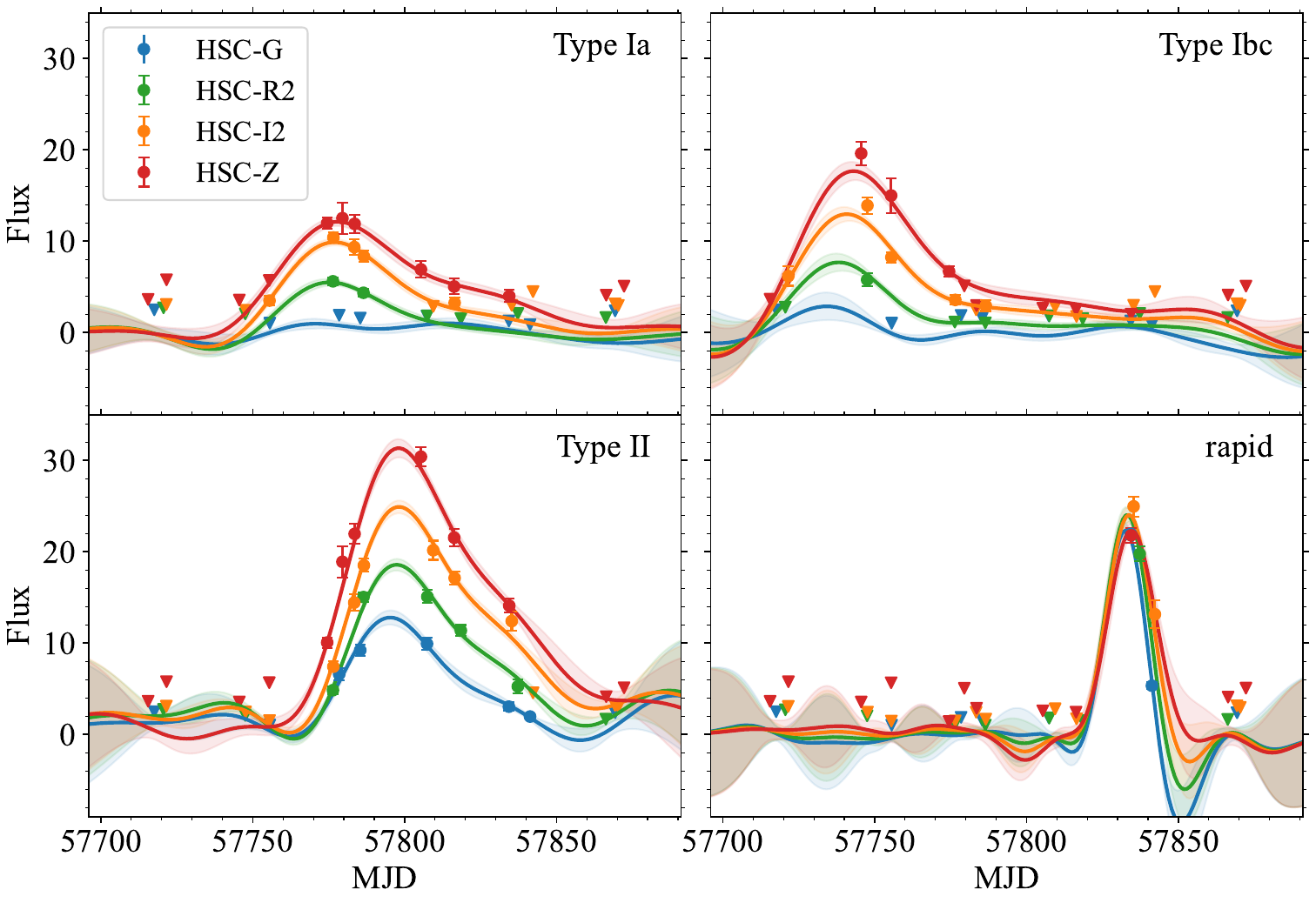}
\caption{Examples of simulated light curves. The observational epoch and depth follow the actual HSC-SSP observations. The circles show the significant detection with $5\sigma$ and thze triangles represent the $3\sigma$ upper limit. Solid lines show the interpolated light curves by 2D~GPR.
 \label{fig:gpr}
  }
\end{center}
\end{figure*}

Our light curves are relatively sparse (cadence of 7-10 days) for measuring the features only from the observed data points. Therefore, we interpolated the light curves by using two-dimensional Gaussian process regression~(2D GPR, ~\citealt{rasmussen2006, pedregosa2011}), a non-parametric supervised learning method with a Python library for machine learning, \texttt{scikit-learn}~\citep{scikit-learn}. We interpolated the light curves in time and wavelength space. The examples of simulated light curves of each SN type~(Type Ia, Ibc, II SNe, and rapid transients) interpolated by 2D GPR are shown in Figure~\ref{fig:gpr}. 

Using these interpolated light curves, we derived various features such as the rise/decline time from the peak to half peak of the light curve, and the flux ratio between different bands. The list of the measured features is shown in Table~\ref{tab:feature}.

\begin{figure}[t]
  \begin{center}
    \includegraphics[scale=0.6]{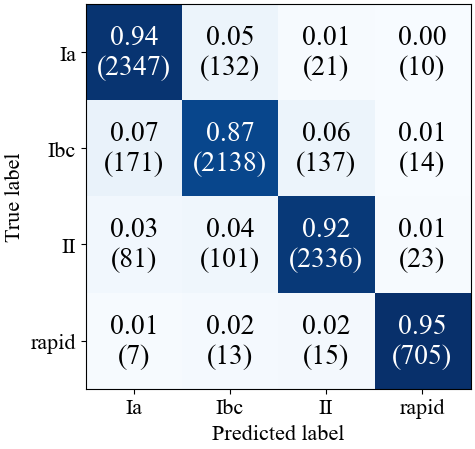}
\caption{Normalized confusion matrix of the RF classifier with the simulation dataset. The numbers in parentheses represent the raw numbers of the objects.
  \label{fig:accuracy}
  }
\end{center}
\end{figure}

With these features derived from the light curves and the redshifts, we developed a RF classifier based on \texttt{scikit-learn}~\citep{scikit-learn}.
Classification accuracy was confirmed by using simulation data which was divided into training data~(75\%) and test data~(25\%). Our RF models classified simulated SNe with an overall accuracy of 93.1\%.
The confusion matrix to show the classification performance for each class is presented in Figure~\ref{fig:accuracy}. 

After checking the classification accuracy, we applied the RF classifier to the real observational data. We processed the observed light curves in the exactly same way with the simulated light curves and derived the same features. 
Then, we classified the observed SN candidates into four types as in the simulated data. As a result, we obtained 315 SN candidates classified as rapid transient.

\subsection{Final sample of rapid transients}
By our RF classifier, a relatively large fraction ($\sim 10\%$) of samples was classified as rapid transient. This is because we used diverse templates covering a wide variety of rapid transients. 
Also, some of our samples suffer from bad data points by imperfect image subtraction during data process. Additionally, some of the light curves  have only partial coverage, which makes the identification as a rapid transient less robust. 

Thus, we imposed two criteria: (1) no bad data points resulting from reduction process, and
(2) at least one observation in any bands before and after the peak of the light curves. Bad data mainly come from the miss-measurement of the flux due to the failure of image subtraction. These criteria removed 240 objects among 315 rapid transient candidates. The fraction of removed objects is rather high as bad data points often have a very high flux, which tends to mimic rapid transients

Finally, we visually checked the light curves of our candidates by comparing their light curves with those of normal Type Ibc SNe~\cite{taddia2015} and literature rapid transient samples~\citep{drout2014, pursiainen2018, hosseinzadeh2018} in the rest frame. We removed the transients whose light curve behavior can be consistent with the error region of the Type Ibc SNe light curve templates by~\citealt{taddia2015}. This corresponds to the time above half max $\Delta t_{1/2}\leq 20~\rm{days}$. These contaminations were classified as rapid transient by the RF classifier because our templates of rapid transients prepared to cover a wide range of the timescale and some of them are similar to typical SNe~(Figure~\ref{fig:phase-sim}). Also, some objects suffer from imperfect interpolation of the light curves by GPR, which erratically resulted in a short timescale. After these quality cut, we obtained 14 rapid transients (Figure \ref{fig:rapid-color}). 

\section{results}
\label{sec:results}
\begin{table*}[t]
    \caption{Overview of the HSC rapid transients}
    \label{tab:rapid_prof}
    \centering
  \begin{tabular}{cccrccccclrc}
  	\hline
	IAU Name & HSC Name & Field~Layer & R.A. & DEC.& $M_{\rm peak}$ & $t_{1/2}$ & Redshift$^{c}$ & Redshift source & Sub group\\
	\hline \hline
	AT 2016jll & HSC16apsu  & COSMOS~UD & 150.48176 & $+$2.43983 & $-18.5$ & 16.1 & 0.741 & \spz & Ibn-like\\
	AT 2017kih & HSC17auls$^{a}$ & COSMOS~UD &149.47972  & $+$2.41892 & $-17.4$ & 11.8 & 0.339 & \spz \\
	AT 2017kii & HSC17bhyl$^{a}$ & COSMOS~UD &150.34255 & +2.03149 & $-18.4$ & 12.2 &0.750  & \spz & Ibn-like\\
	AT 2017kij & HSC17btum$^{a}$ & COSMOS~UD & 149.47509 & +2.66588 & $-17.6$ & 15.4 & $0.467^{+0.010}_{-0.011} $ & COSMOS \phz \\
	AT 2017kil & HSC17dadp$^{a}$ & COSMOS~UD & 150.12104 & +1.60768 & $-18.6$ & 12.0 &$0.830^{+0.055}_{-0.047}$ &  HSC UD \phz  \\
    SN 2017fei & HSC17dbpf$^{b}$ & COSMOS~UD & 149.63925 & +1.99158 & $-20.8$ &12.8 & 1.851 & spec-$z$ & SLSN-like\\
	AT 2017kim & HSC17dhvg & COSMOS~UD & 149.87828 & +1.56411 & $-17.1$ & 10.6 &$0.730^{+0.036}_{-0.025}$ & HSC UD \phz \\
	AT 2017kin & HSC17dodm & COSMOS~UD & 150.79599 & +2.16747 &$-18.6$ & 7.0 &1.476 & \spz   \\
      	\hline	
	AT 2017kik & HSC17cgdi  & COSMOS~D &151.02149 & +2.84606 & $-18.6$ & 12.1 & $0.805$ & spec-$z$ \\
	\hline
	AT 2019aavm & HSC19aqfi & SXDS~UD & 34.80293 & $-$4.93046 & $-17.9$& 13.7 & 0.851 & SPLASH \phz \\
	AT 2019aavp & HSC19bfbh & SXDS~UD & 34.31801 & $-$4.93499 & $-19.0$& 10.0 & 1.481  & SPLASH \phz \\
	AT 2019aavq & HSC19bijv & SXDS~UD & 34.37900 & $-$5.49328 & $-19.1$ & 9.8 & 1.486 & SPLASH \phz \\
	\hline
    AT 2019aavn & HSC19bbhh & SXDS~D & 35.68173 & $-$4.05113 & $-22.0$& 17.0 & $1.370^{+0.051}_{-0.049} $ & HSC \phz & SLSN-like\\
	AT 2019aavo & HSC19beav & SXDS~D & 35.96370& $-$5.80319 & $-17.7$& 18.6 & $0.312 $ & \spz \\
	\hline
  \end{tabular}
   \tablecomments{
   $^a$ reported by \citealt{tampo2020}, $^b$ reported by \citealt{moriya2019}, $^{c}$ Redshifts of the putative host galaxies
   }
\end{table*}

\begin{figure*}[t]
\begin{minipage}[b]{0.33\linewidth}
    \centering
    \includegraphics[width=1.0\linewidth]{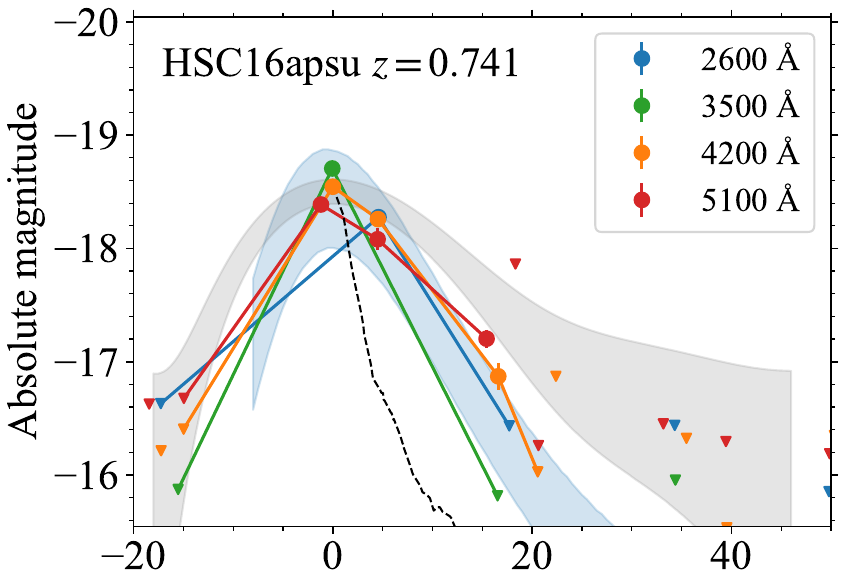}
  \end{minipage}  
  \begin{minipage}[b]{0.33\linewidth}
    \centering
    \includegraphics[width=1.0\linewidth]{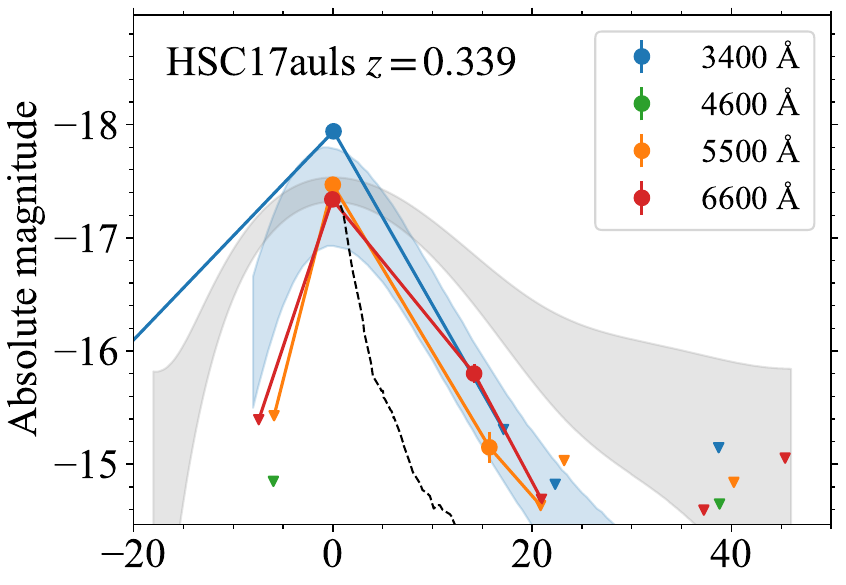}
  \end{minipage}
    \begin{minipage}[b]{0.33\linewidth}
    \centering
    \includegraphics[width=1.0\linewidth]{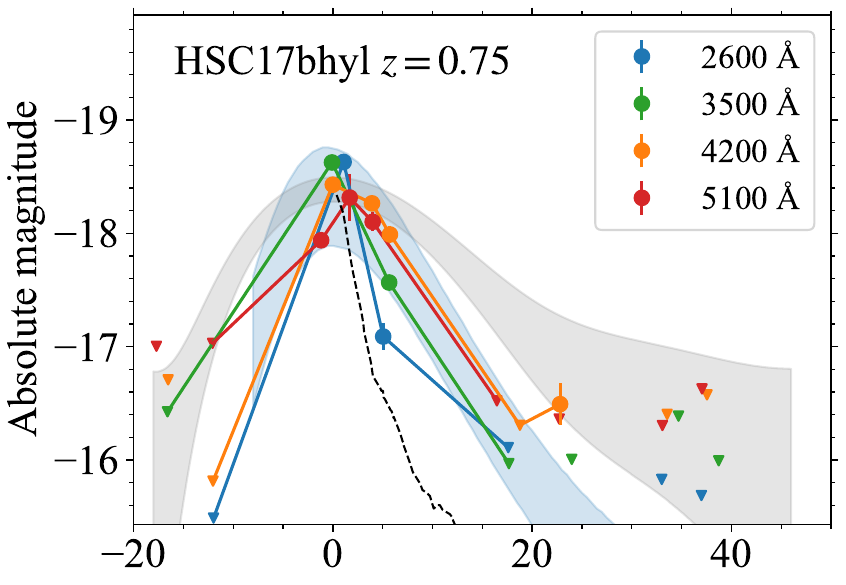}
  \end{minipage}
  \begin{minipage}[b]{0.33\linewidth}
    \centering
    \includegraphics[width=1.0\linewidth]{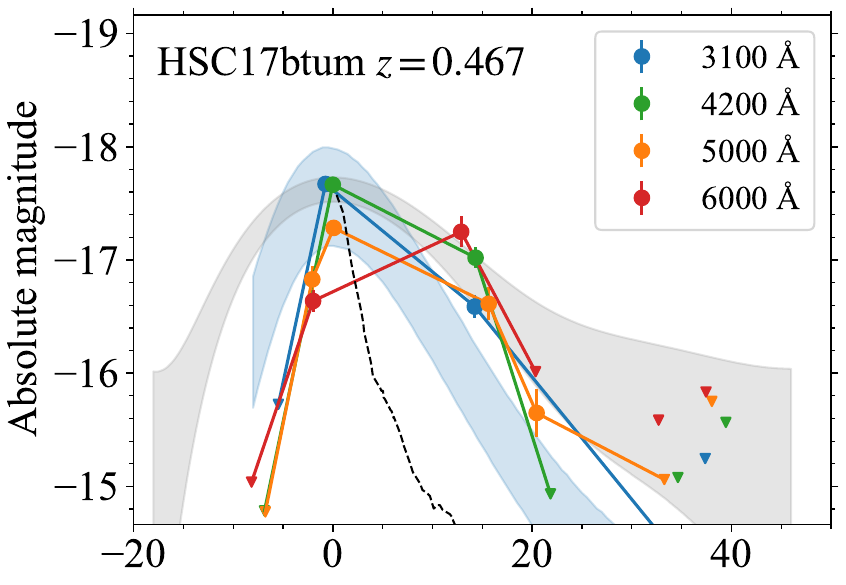}
  \end{minipage}
    \begin{minipage}[b]{0.33\linewidth}
    \centering
    \includegraphics[width=1.0\linewidth]{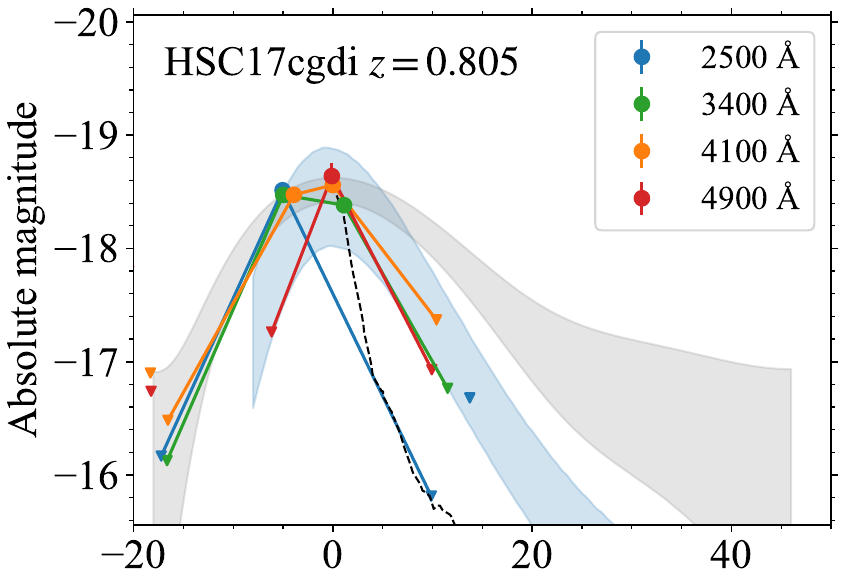}
  \end{minipage}
\begin{minipage}[b]{0.33\linewidth}
    \centering
    \includegraphics[width=1.0\linewidth]{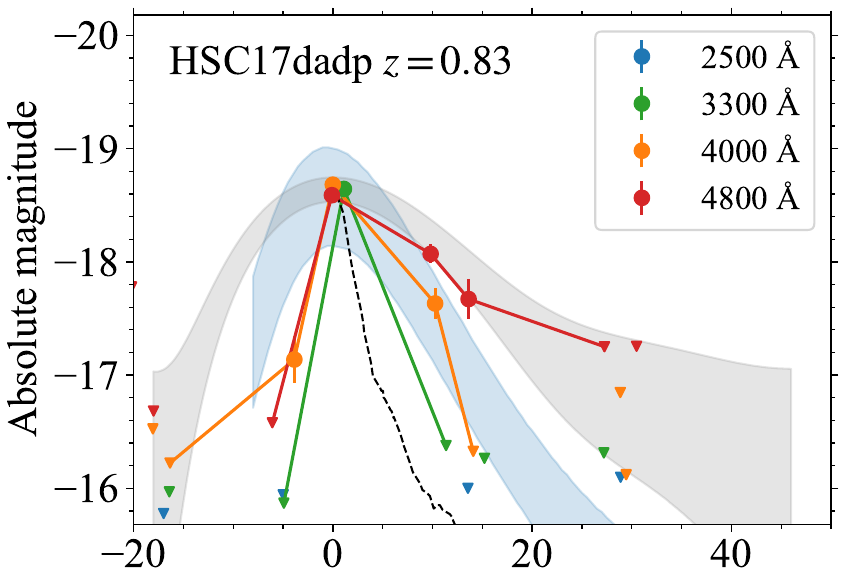}
  \end{minipage}  
    \begin{minipage}[b]{0.33\linewidth}
    \centering
    \includegraphics[width=1.0\linewidth]{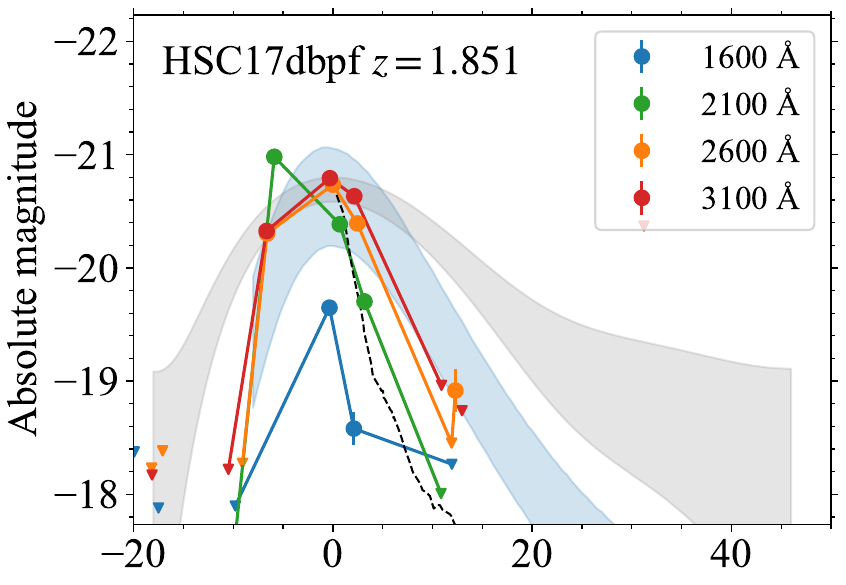}
  \end{minipage}
  \begin{minipage}[b]{0.33\linewidth}
    \centering
    \includegraphics[width=1.0\linewidth]{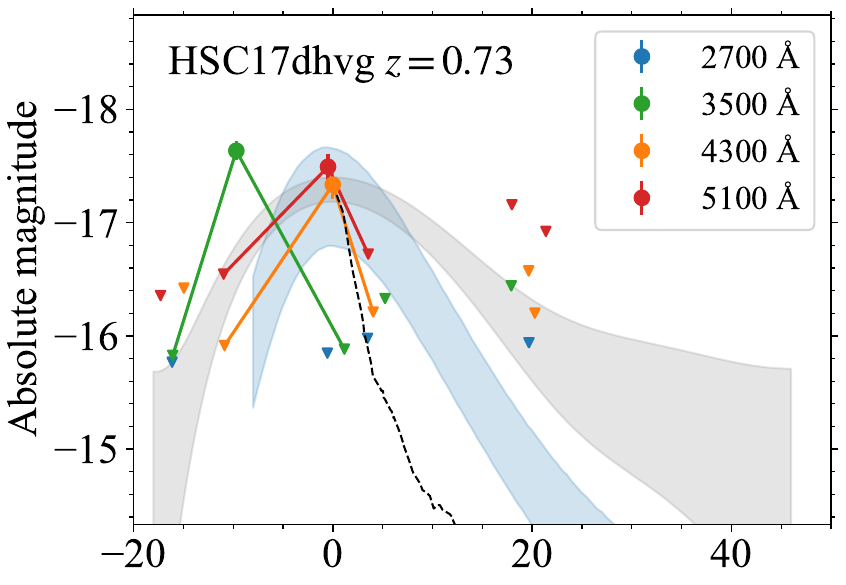}
  \end{minipage}
    \begin{minipage}[b]{0.33\linewidth}
    \centering
    \includegraphics[width=1.0\linewidth]{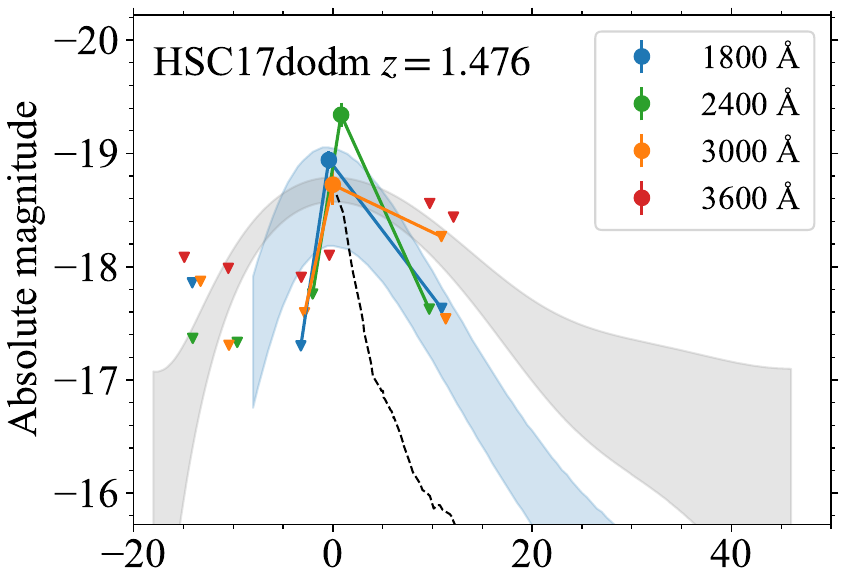}
  \end{minipage}
\begin{minipage}[b]{0.33\linewidth}
    \centering
    \includegraphics[width=1.0\linewidth]{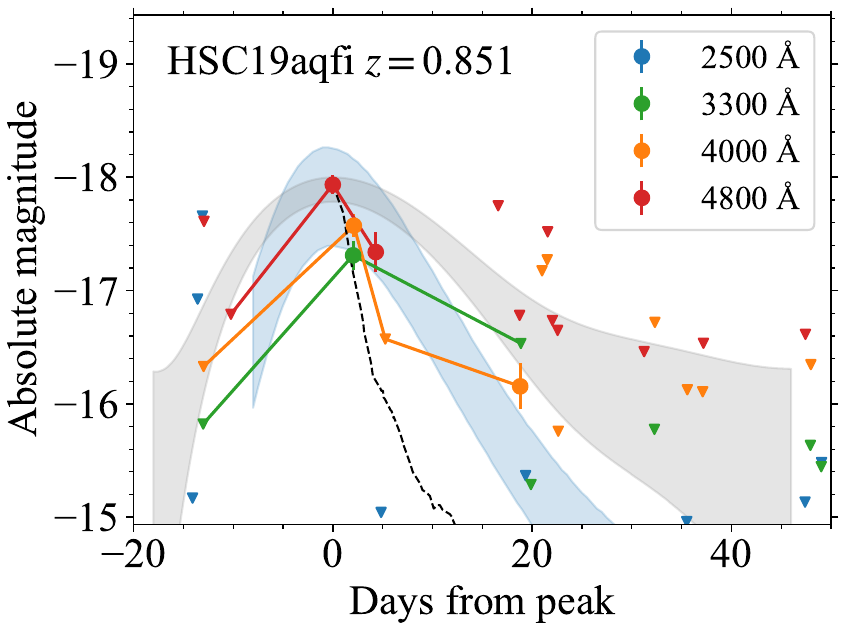}
  \end{minipage}  
  \begin{minipage}[b]{0.33\linewidth}
    \centering
    \includegraphics[width=1.0\linewidth]{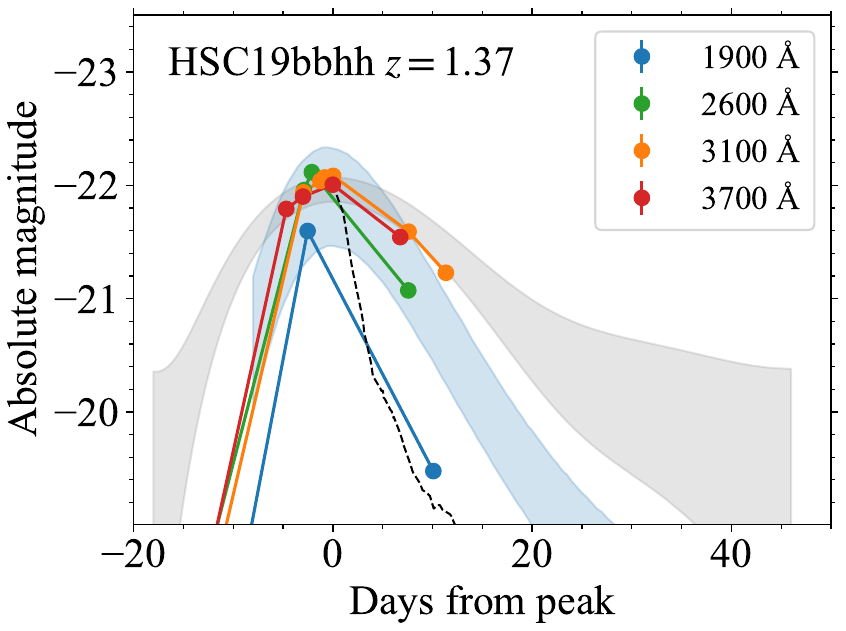}
  \end{minipage}
      \begin{minipage}[b]{0.33\linewidth}
    \centering
    \includegraphics[width=1.0\linewidth]{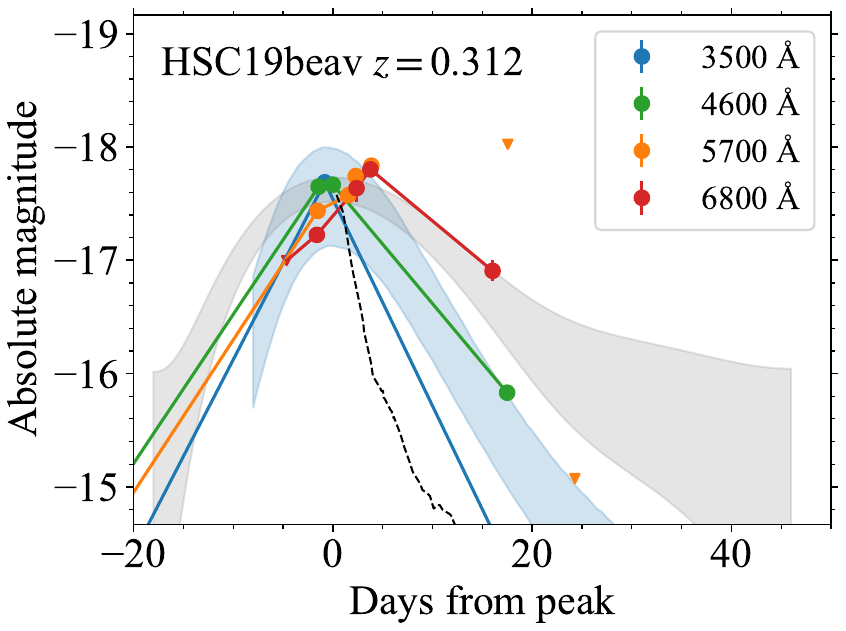}
  \end{minipage}
    \begin{minipage}[b]{0.5\linewidth}
    \centering
    \includegraphics[width=0.66\linewidth]{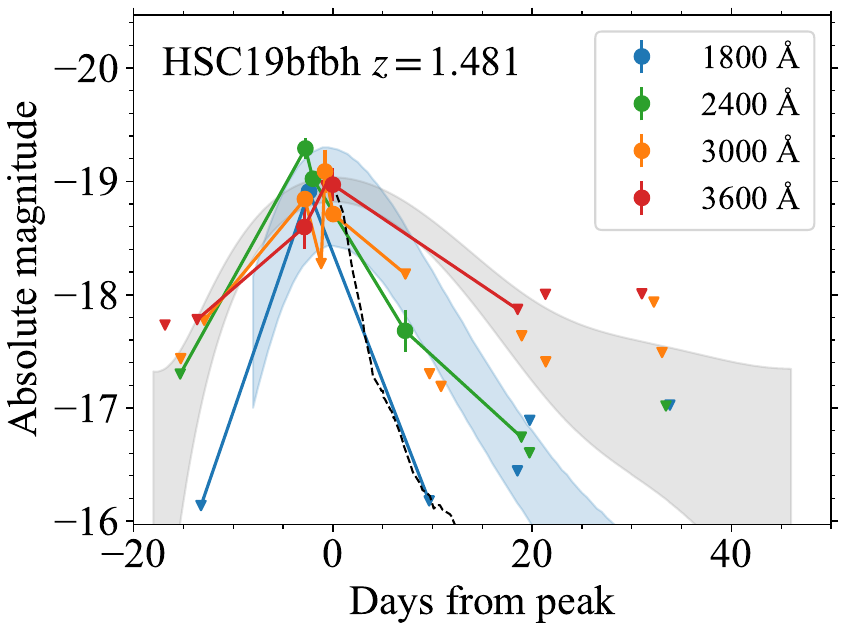}
  \end{minipage}
      \begin{minipage}[b]{0.5\linewidth}
    \centering
    \includegraphics[width=0.66\linewidth]{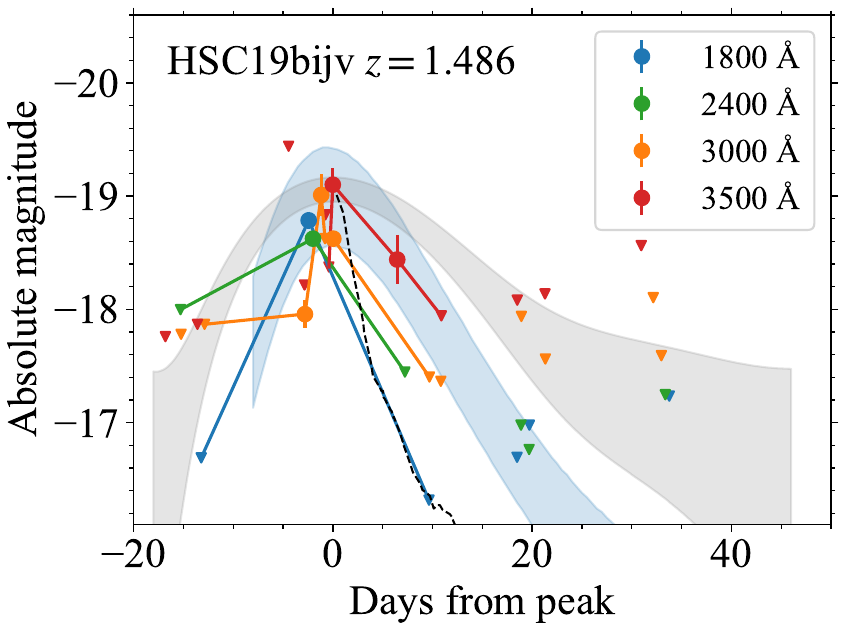}
  \end{minipage}
  \caption{
  \label{fig:rapid}
  The light curves of HSC rapid transient samples. To define the light curve peak, we use a band corresponding to the rest frame $g$-band~($z<0.2$:~HSC-G, $0.2<z<0.5$:~HSC-R2, $0.5<z<0.8$:~HSC-I2, $0.8<z$:~HSC-Z)~and set the peak date in the band as $t=0$. The gray shaded region shows light curve template of Type Ibc SNe in $g$-band from \cite{taddia2015} while the blue shaded region shows light curve template of Type Ibn SNe with error in $R$-band from~\cite{hosseinzadeh2018}. The black dashed line is the light curve of AT2018cow in $g$-band as the representative of the LFBOTs. 
   }
\end{figure*}

In this section, we give an overview of the rapid transients identified in our work. Figure~\ref{fig:rapid} shows the light curves of our samples in the rest frame. We define the light curve peak in the band corresponding to the rest frame $g$-band~($z < 0.2$: HSC-G, $0.2 < z < 0.5$: HSC-R2, $0.5 < z < 0.8$: HSC-I2, $0.8 < z$: HSC-Z). Although the objects at high redshift ($z > 1.0$) do not have the band corresponding to the rest frame $g$-band, we use HSC-Z band as the closest band to the rest frame $g$-band. For absolute magnitude in Figure~\ref{fig:rapid}, we apply a simple $K$ correction~($M=m-\mu+2.5\log{(1+z)}$) only by considering the term of the cosmic expansion as the intrinsic SEDs of these transients are not well established. To compare the timescale of our samples, we also show the light curve templates of Type Ibc SNe ~(gray;~\citealt{taddia2015}) and Type Ibn SNe light curve (blue; \citealt{hosseinzadeh2018}) as well as the light curve of AT 2018cow~\citep{perley2019} as a representative of LFBOTs.

\begin{figure*}[t]
\begin{minipage}[b]{0.33\linewidth}
    \centering
    \includegraphics[width=1.0\linewidth]{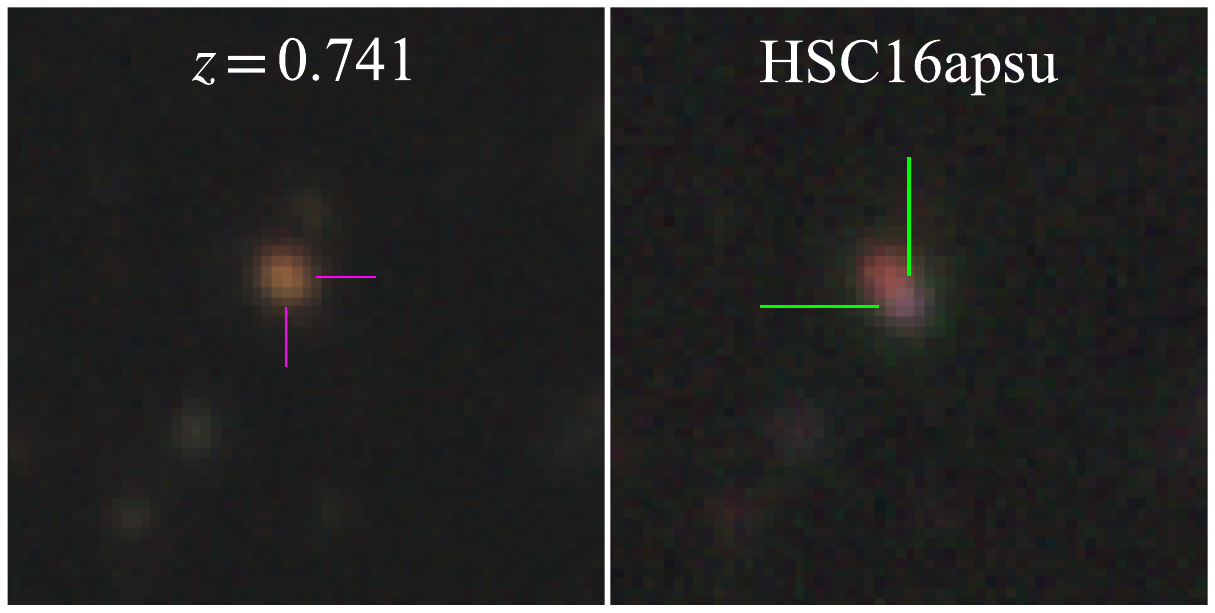}
  \end{minipage}  
  \begin{minipage}[b]{0.33\linewidth}
    \centering
    \includegraphics[width=1.0\linewidth]{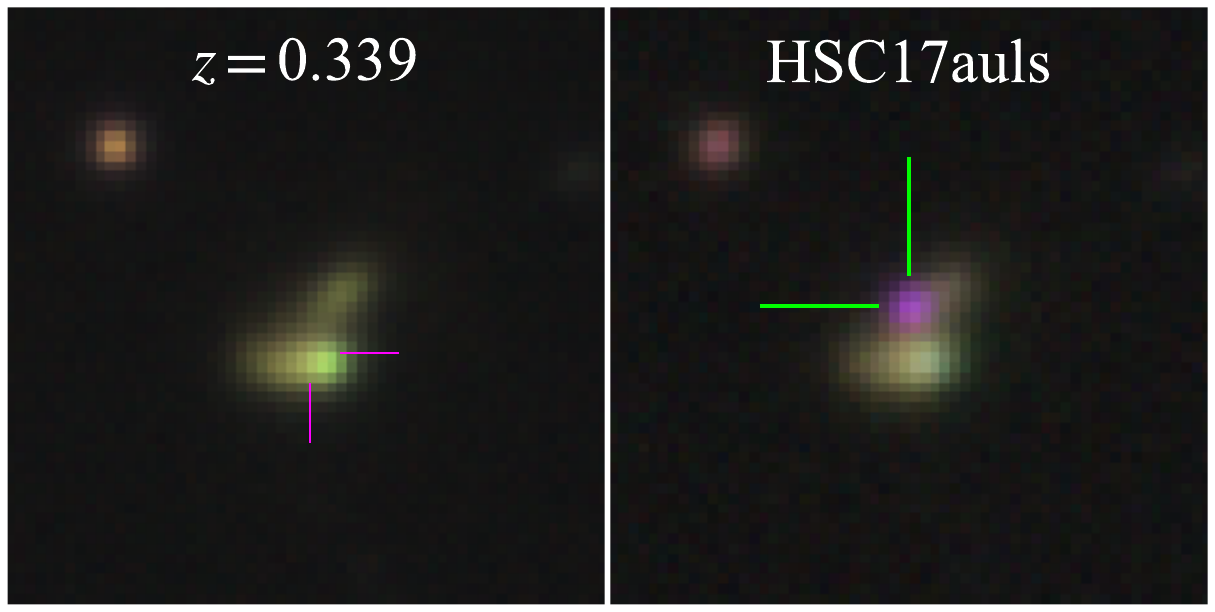}\\
  \end{minipage}
    \begin{minipage}[b]{0.33\linewidth}
    \centering
    \includegraphics[width=1.0\linewidth]{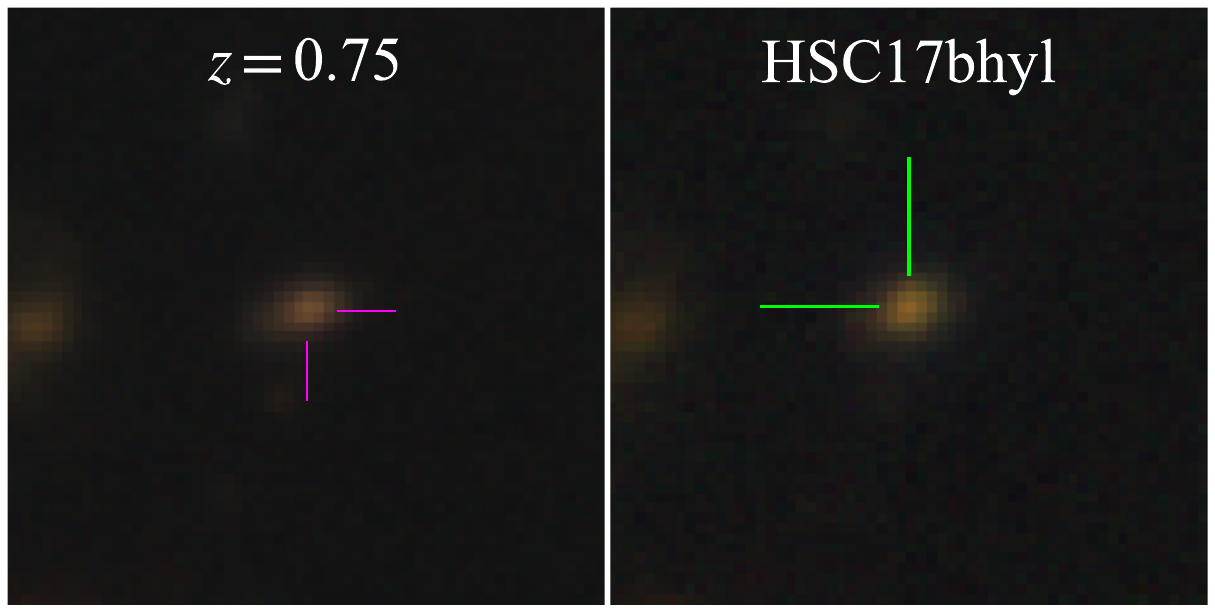}
  \end{minipage}
  \begin{minipage}[b]{0.33\linewidth}
    \centering
    \includegraphics[width=1.0\linewidth]{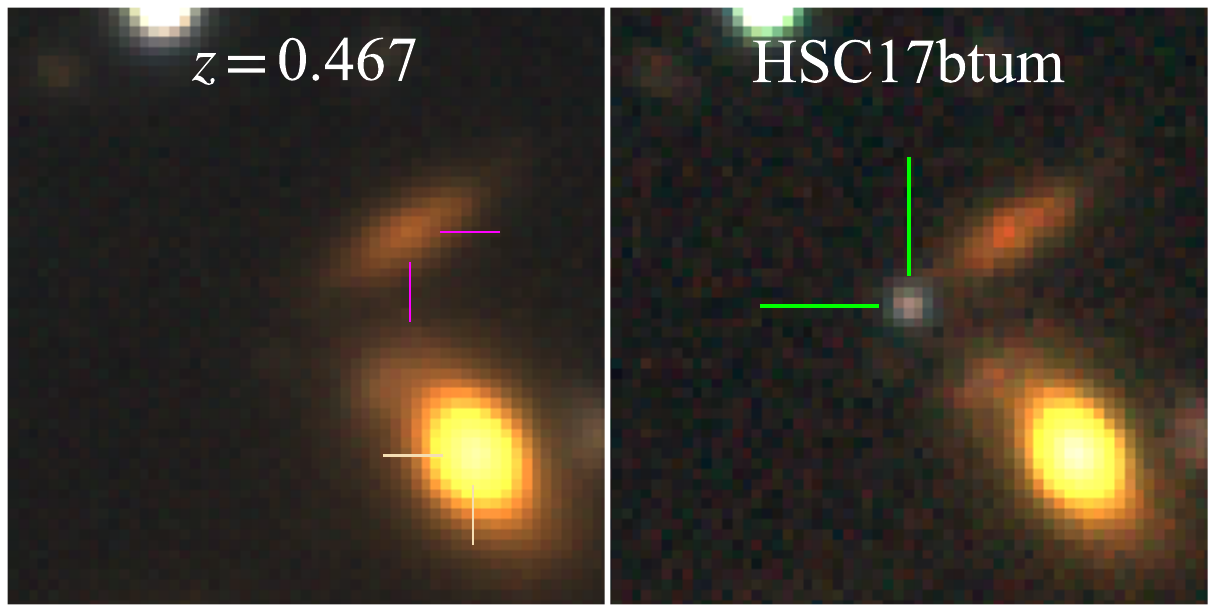}
  \end{minipage}
\begin{minipage}[b]{0.33\linewidth}
    \centering
    \includegraphics[width=1.0\linewidth]{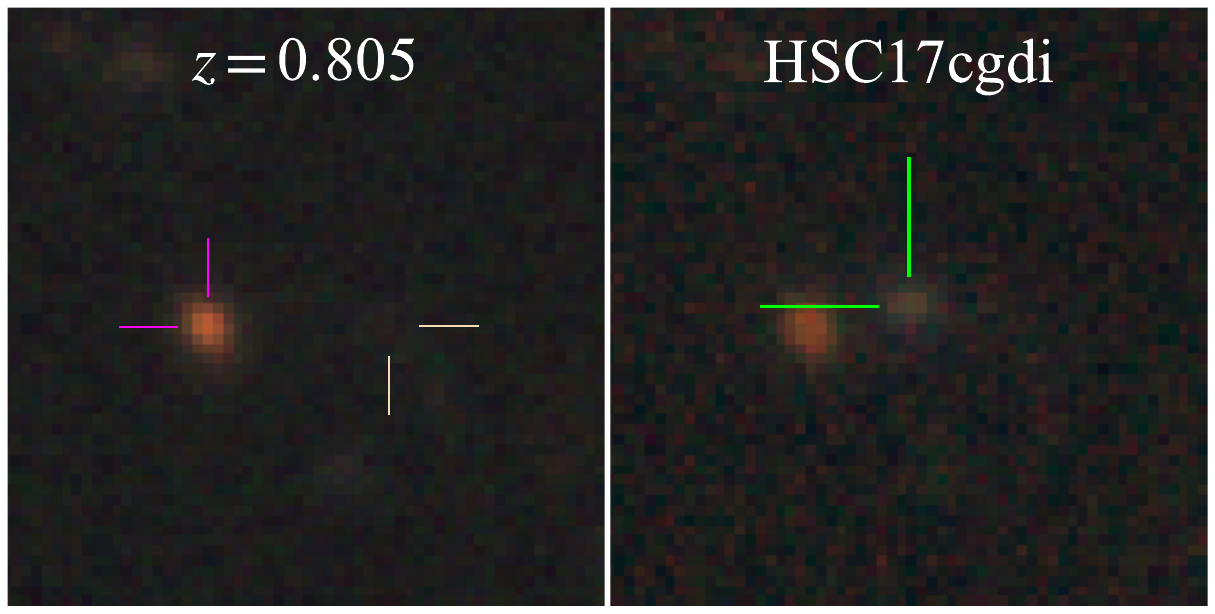}
  \end{minipage}  
  \begin{minipage}[b]{0.33\linewidth}
    \centering
    \includegraphics[width=1.0\linewidth]{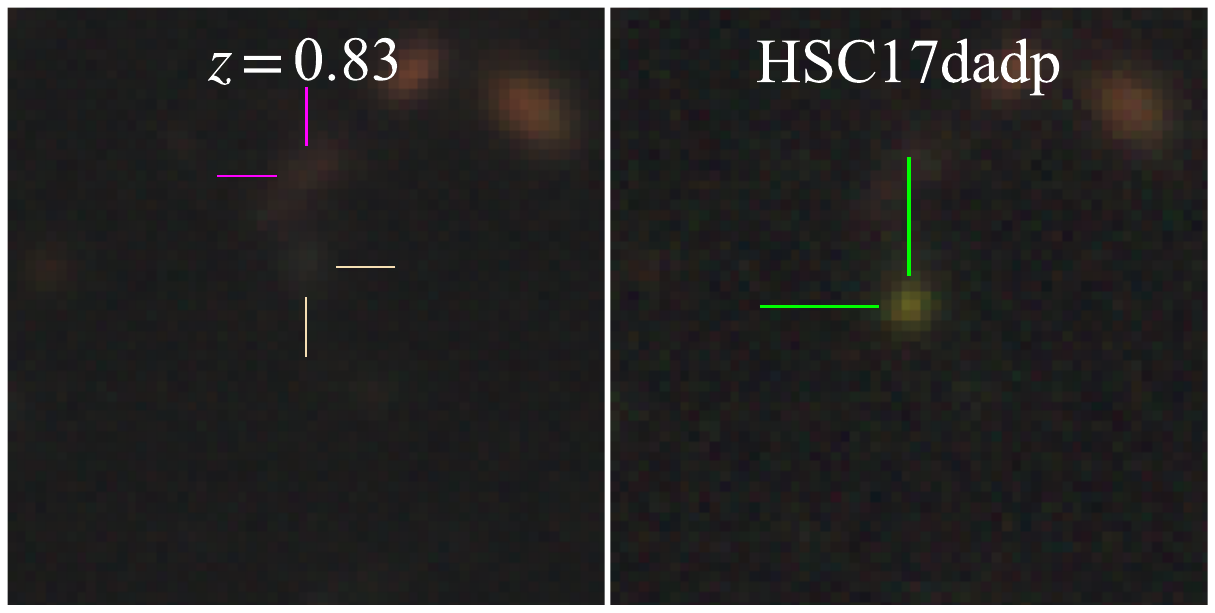}
  \end{minipage}
    \begin{minipage}[b]{0.33\linewidth}
    \centering
    \includegraphics[width=1.0\linewidth]{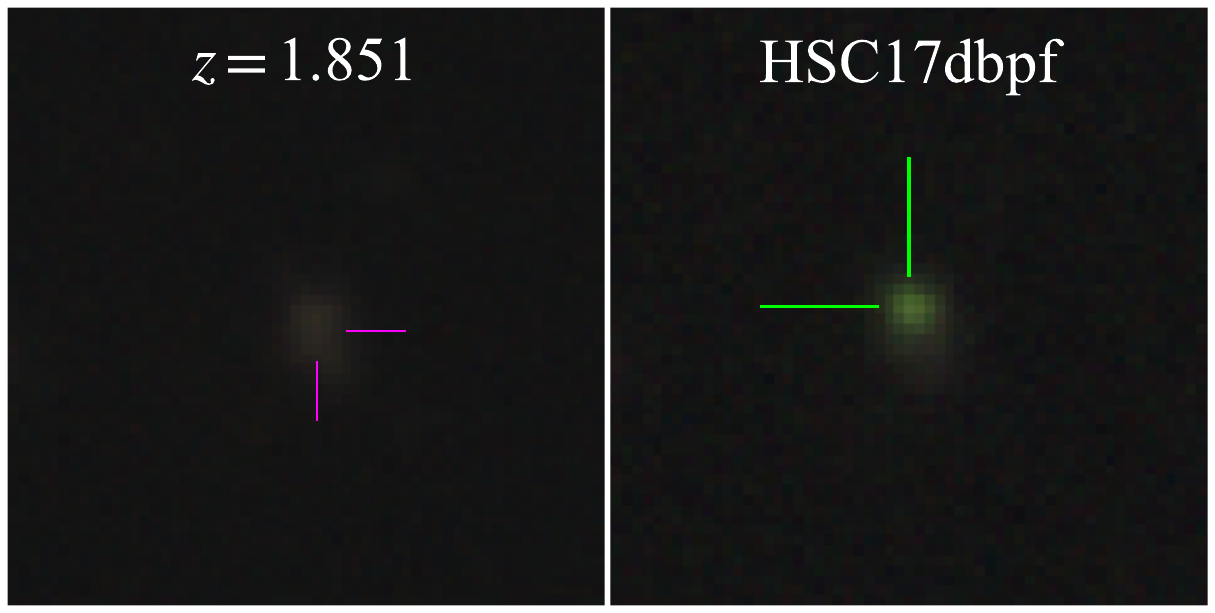}
  \end{minipage}
  \begin{minipage}[b]{0.33\linewidth}
    \centering
    \includegraphics[width=1.0\linewidth]{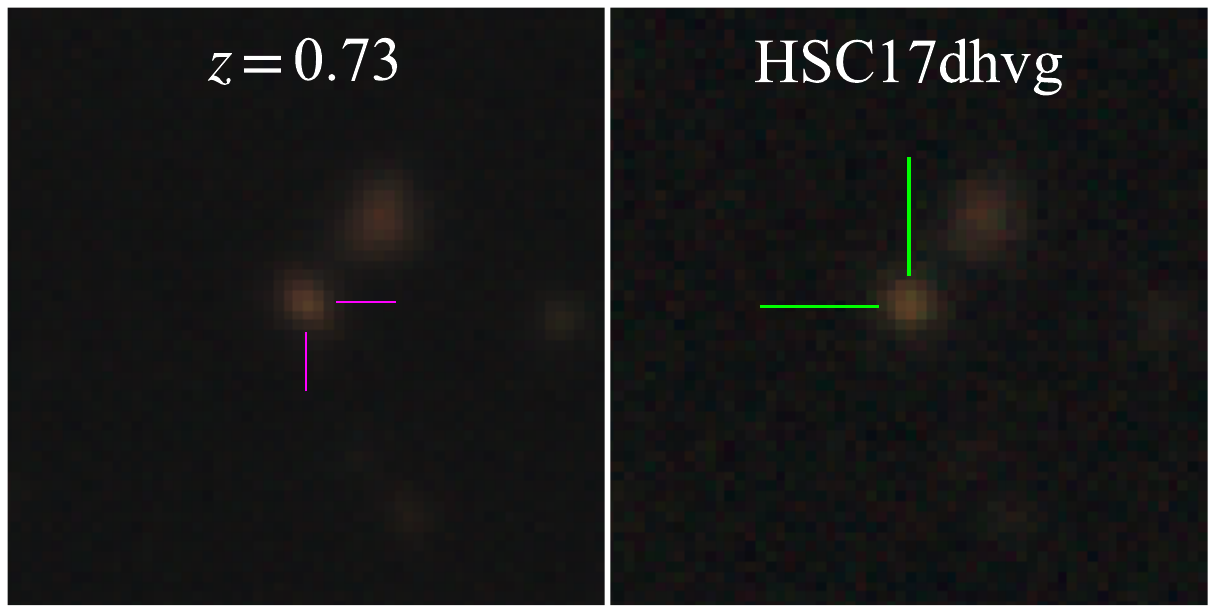}
  \end{minipage}
      \begin{minipage}[b]{0.33\linewidth}
    \centering
    \includegraphics[width=1.0\linewidth]{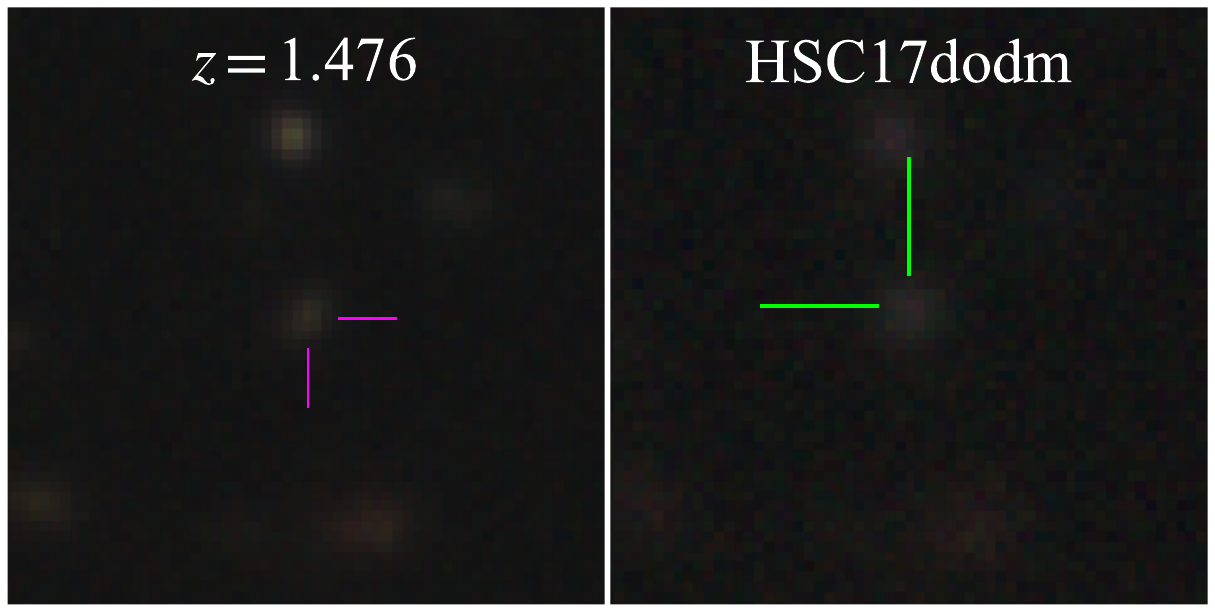}
  \end{minipage}
  \begin{minipage}[b]{0.33\linewidth}
    \centering
    \includegraphics[width=1.0\linewidth]{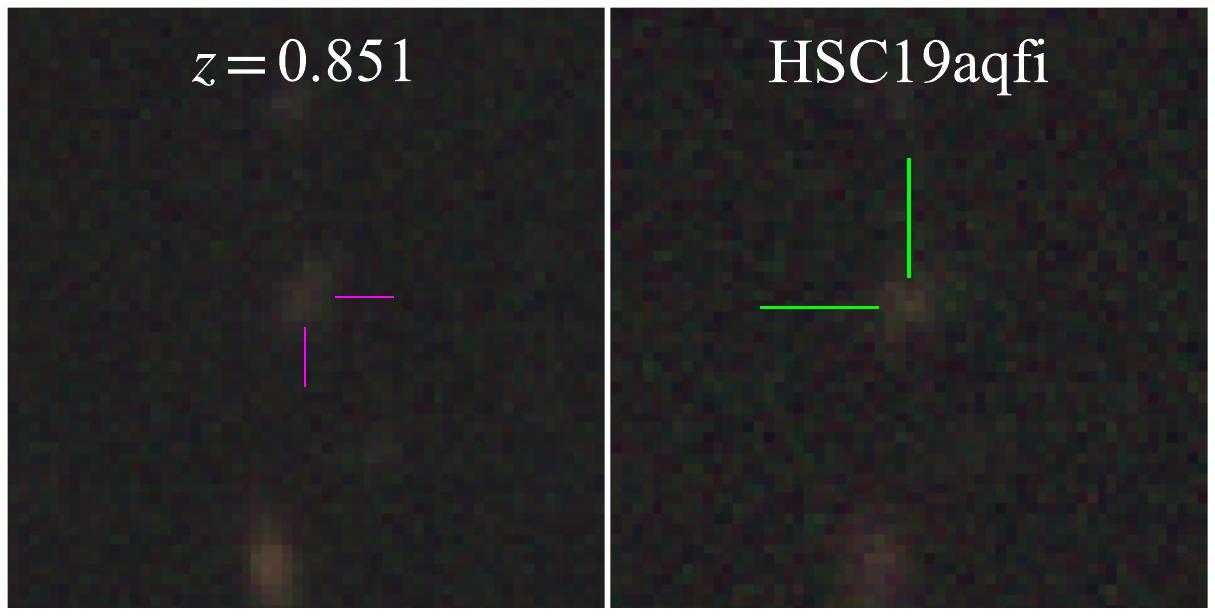}
  \end{minipage}
      \begin{minipage}[b]{0.33\linewidth}
    \centering
    \includegraphics[width=1.0\linewidth]{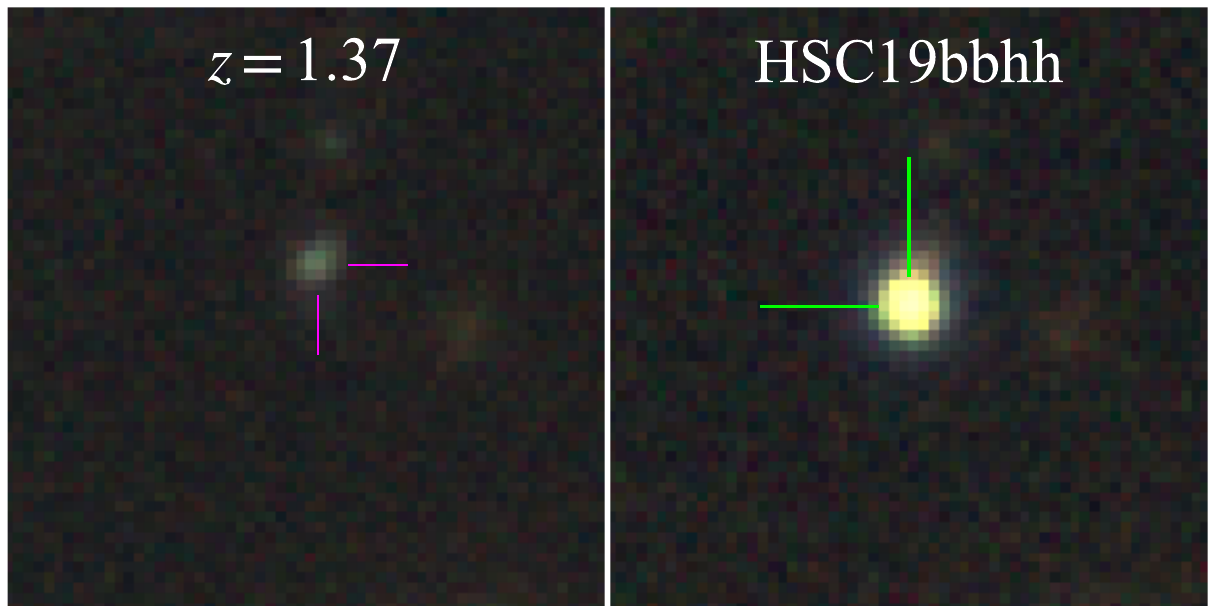}
  \end{minipage}
  \begin{minipage}[b]{0.33\linewidth}
    \centering
    \includegraphics[width=1.0\linewidth]{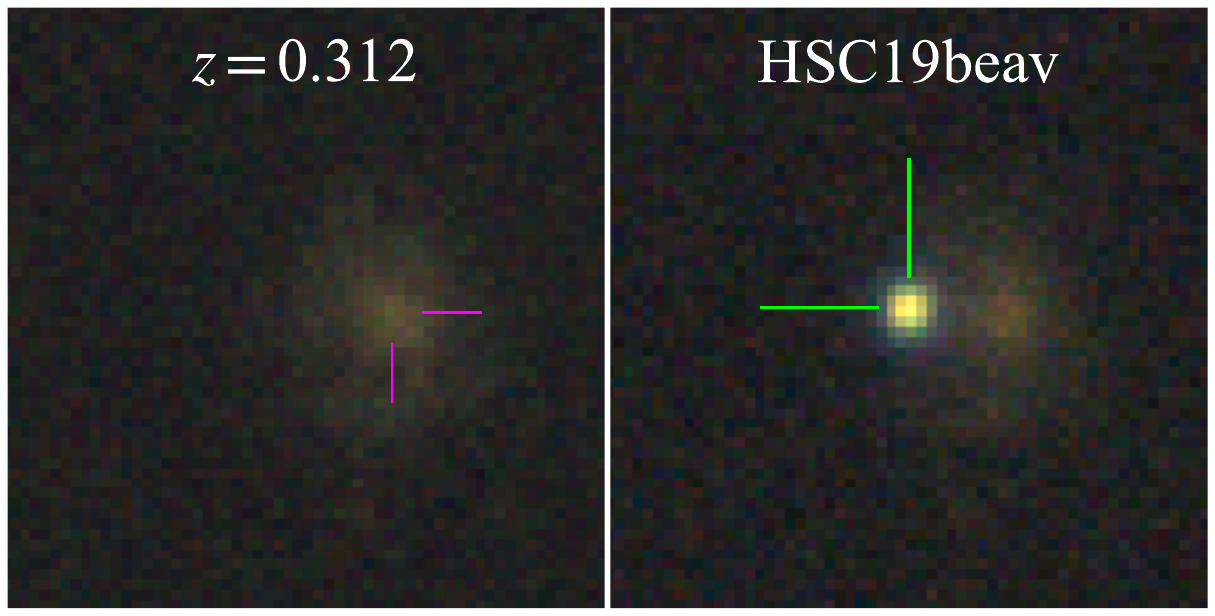}
  \end{minipage}
  \begin{minipage}[b]{0.5\linewidth}
    \centering
    \includegraphics[width=0.66\linewidth]{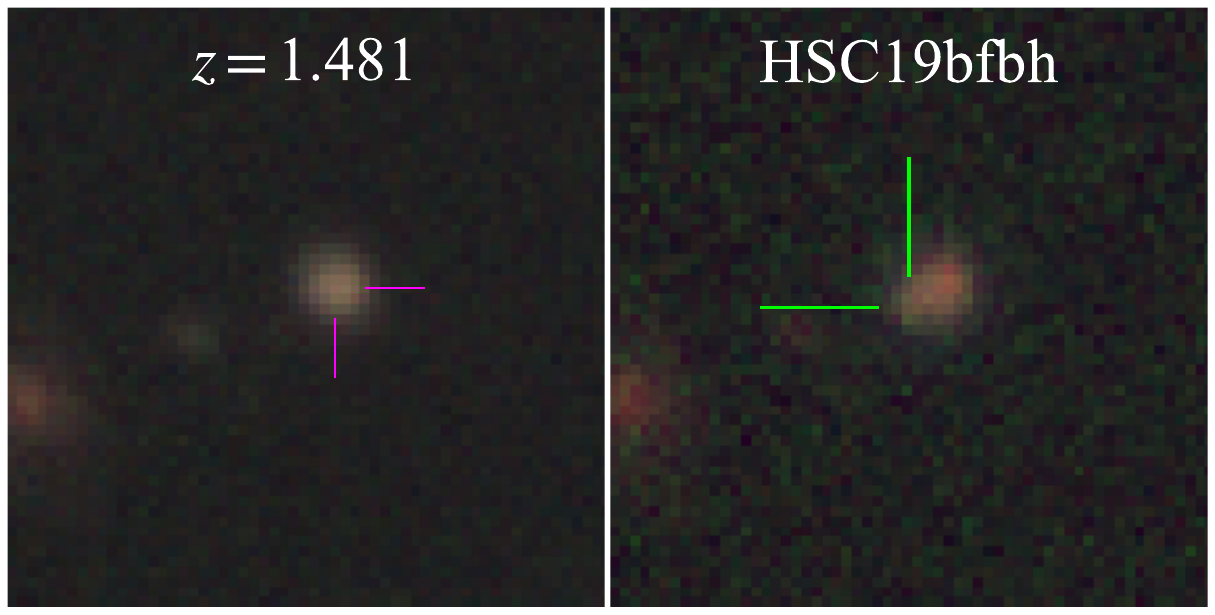}
  \end{minipage}
  \begin{minipage}[b]{0.5\linewidth}
     \centering
    \includegraphics[width=0.66\linewidth]{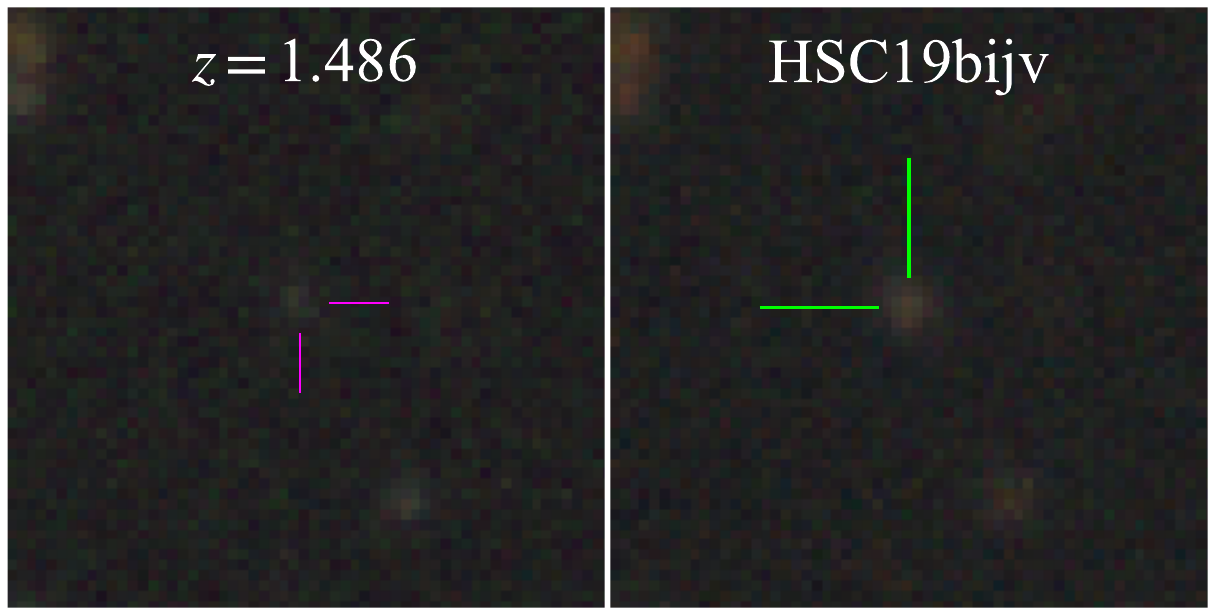}
  \end{minipage}
    \caption{Three-color images of the rapid transients (green) and their putative host galaxies (magenta). The left panels show the reference images, and the right panels show the images after the SN discovery at around the peak brightness. Each image has a size of $10''\times10''$ centered at the SN location. North is up and east is left. We also marked the other marginal candidates of host galaxies (white) for HSC17btum, HSC17cgdi,and  HSC17dadp.
  \label{fig:rapid-color}
   }
  \end{figure*}

In Table~\ref{tab:rapid_prof}, we summarize basic information of our samples. To further check the reliability of the host galaxy identification, we calculate normalized distance $d/r$ to surrounding galaxies by following \cite{gupta2016}. Here $d$ is offset from the galaxy center and $r$ is the radius of the galaxy. Then we assign host galaxies as those with the smallest $d/r$.
We confirmed that the results using the normalized distance are basically consistent with the identifications by \cite{yasuda2019}. There are only two exceptions: HSC17cgdi and HSC17dadp. For these two objects, the galaxies with the smallest normalized distance are very faint and have very high photometric redshifts~(1.72 and 2.78, respectively). If we assume these redshifts for these transients, the detection in the bluest band corresponds to 1600 \AA~in the rest frame, which is not likely to happen as it causes an extremely high luminosity in short UV. Therefore, we consider the second closest galaxies in normalized distance as host galaxies, which are those identified by \cite{yasuda2019}. For the case of HSC17btum, we confirmed that the host galaxy identified by \cite{yasuda2019} is closest in normalized distance $d/r$. Although there is the second closest galaxy in similar offset (Figure \ref{fig:rapid-color}), the redshift of this object is estimated as $z=0.43$ which is similar to the redshift of the closest galaxy ($z=0.467$). Therefore, even if the second closest galaxy is the host, it would not significantly affect the luminosity or timescale of the HSC17btum.

\begin{figure}[t]
  \begin{center}
    \includegraphics[scale=0.45]{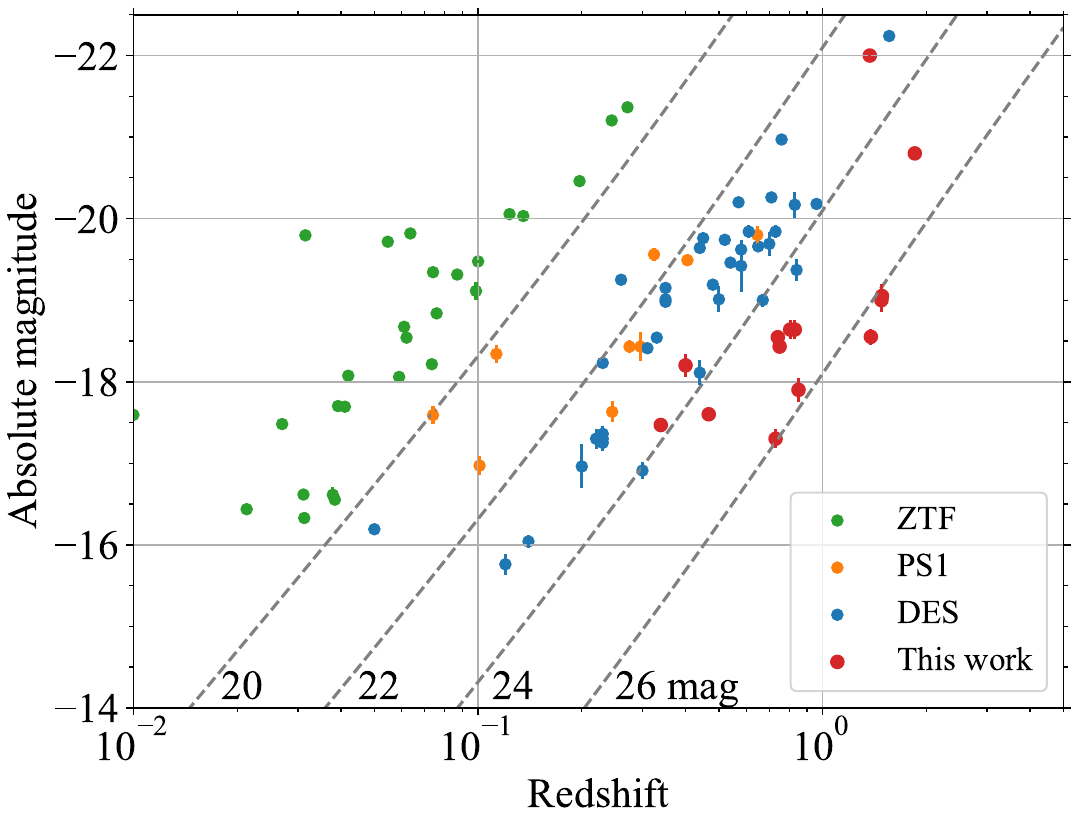}
\caption{Redshift distribution of our samples. We also plot the rapid transient samples from ZTF~\citep{ho2023}, PS1~\citep{drout2014}, and DES~\citep{pursiainen2018}. The dashed line represents detection limits for each limiting magnitude.
  \label{fig:ZvsM}
  }
\end{center}
\end{figure}

For seven objects among our samples~(HSC16apsu, HSC17auls, HSC17bhyl, HSC17cgdi, HSC17dbpf, HSC17dodm, HSC19beav), redshifts are measured by spectroscopic observations of their putative host galaxies. For the other objects, the redshifts are estimated by multicolor photometric data: COSMOS2015 catalog~\citep{laigle2016} for HSC17btum, SPLASH photometric redshift catalog~\citep{mehta2018} for HSC19aqfi, HSC19bfbh, HSC19bijv, and photometric redshift from the HSC-SSP survey~\citep{tanakamasayuki2018} for HSC17dadp, HSC17dhvg, HSC19bbhh. We show the redshift distribution of our samples in Figure~\ref{fig:ZvsM}, as compared with the samples from ZTF~\citep{ho2023}, PS1~\citep{drout2014}, and DES~\citep{pursiainen2018}. Our samples cover a medium to high redshift range~($0.34\leq z\leq 1.85$).

\begin{figure*}[t]
  \begin{center}
    \includegraphics[scale=0.95]{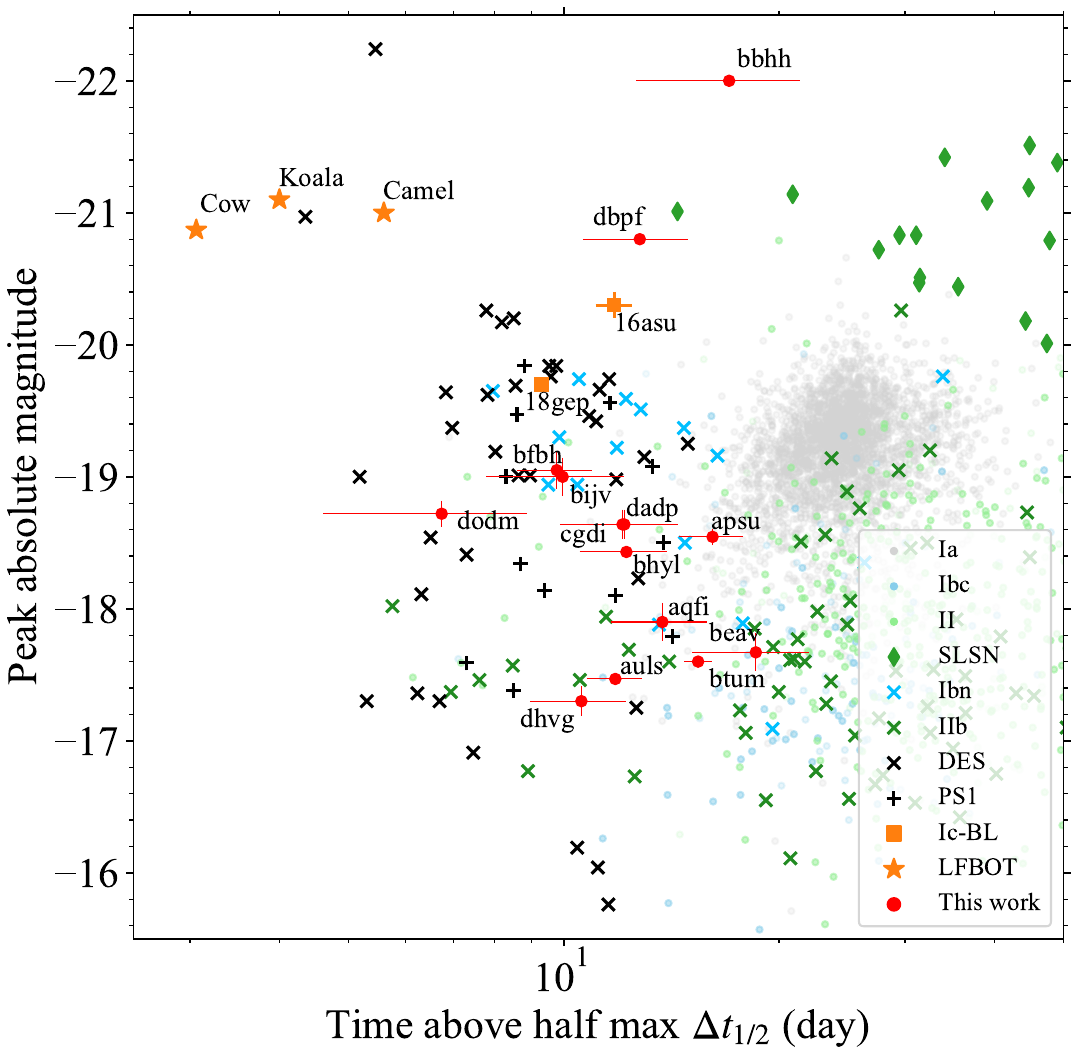}
\caption{Phase diagram of our samples~(red points) with literature SNe~\citep{perley2020, pursiainen2018, drout2014}. Time above half max time~$\Delta t_{1/2}$ and the peak magnitude in rest frame $g$-band are plotted. The red points are rapid transients from the Subaru HSC-SSP transient survey.
 \label{fig:phase}
  }
\end{center}
\end{figure*}

Our samples are shown in the luminosity-duration parameter space with literature SN samples~(Figure~\ref{fig:phase}). We define the rise time~($t_{1/2,\rm{rise}}$) and the decline time~($t_{1/2,\rm{decline}}$) as the time between the peak epoch and the epochs with the flux half of the peak in the interpolated light curves. Most of our samples are located at a similar parameter space with Type Ibn or Type IIb SNe. Although two samples~(HSC17dbpf, HSC19bbhh) are as luminous as LFBOTs such as AT2018cow, light curves of these two transients evolve more slowly than known LFBOTs. In fact, HSC17dbpf and HSC19bbhh are more similar to unusual Type Ic-BL SNe such as SN2018gep~\citep{ho2019} and iPTF16asu~\citep{whitesides2017, wang2019} or rapidly rising and luminous transients~\citep{arcavi2016}, which are located between SLSNe and LFBOTs in the luminosity-duration parameter space.

\begin{figure}[t]
  \begin{center}
    \includegraphics[scale=0.5]{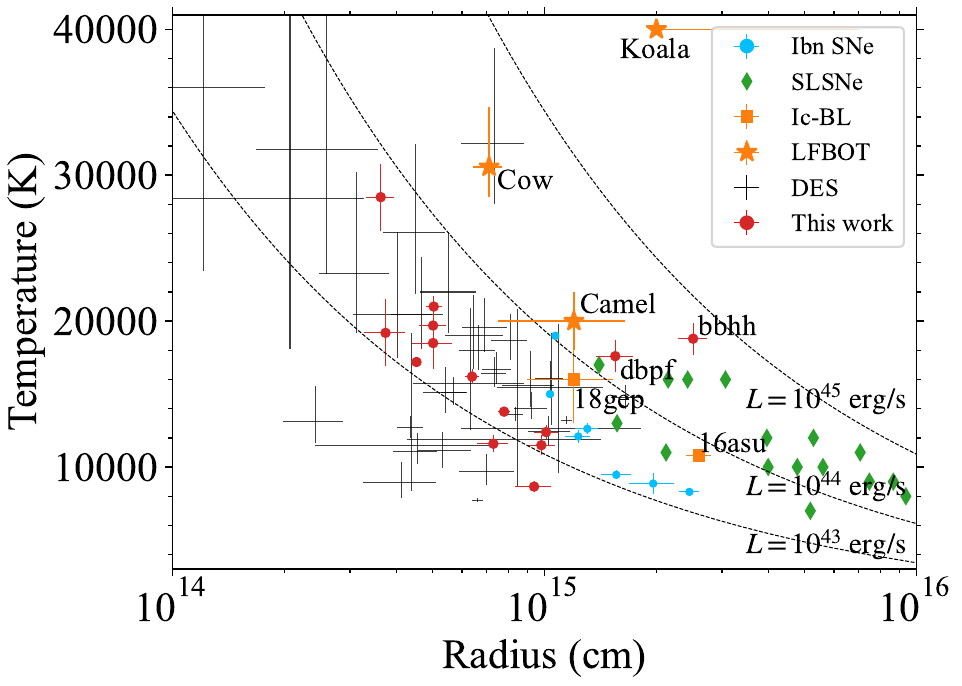}
\caption{Temperture vs. photospheric radius estimated by blackbody fitting. We also plot the rapid transient samples from DES~\citep{pursiainen2018} with black crosses, Type Ibn SNe~\citep{benami2023, karamehmetoglu2021} with blue circles, LFBOTs with orange stars~\citep{perley2019, perley2021, jiang2022}, and rapidly evolving SLSNe with green circles~\citep{whitesides2017, ho2019}. Green crosses represent the normal SLSN from~\citep{zhao2020} for comparisons.
  \label{fig:TvsR}
  }
\end{center}
\end{figure}

We fit the observational data of our samples with blackbody function to estimate the photospheric temperatures and radii at their peak epochs. Here the peak epoch is defined in the band corresponding to rest frame $g$-band. Figure~\ref{fig:TvsR} shows the relation between the temperature and photospheric radius of our samples at the peak epoch. The temperatures range from 9,000 to 29,000~K with a median of 17,000~K and the photospheric radii range from $3.7\times10^{14}$ to $2.5\times 10^{15}~\rm{cm}$ with a median of $7\times10^{14}~\rm{cm}$. 
Most of our samples show temperature and radius 
similar to those of the rapid transients from DES~\citep{pursiainen2018}. On the other hand, two luminous samples~(HSC17dbpf, HSC19bbhh) are located at relatively large radii~($\geq10^{15}~\rm{cm}$) and high temperature~($\geq 15,000~\rm{K}$) parameter region. Although they are not so hot as LFBOTs, their temperatures are higher than those of normal SLSNe~\citep{zhao2020}.

\section{discussion}
\label{sec:discussion}
\subsection{Host galaxies}

Properties of host galaxies also give us insight into the progenitor of rapid transients. In this section, we study the stellar masses and star formation rates of the host galaxies.
For this purpose, we use photometric SED fitting software \texttt{CIGALE}~(Code Investigating GALaxy Evolution;~\cite{boquien2019}).
For the modeling parameters, we follow \cite{liang2024} which analysed the host galaxies of SNe with \texttt{CIGALE} based on the HSC photometry.
Namely, we assume a star formation history with two decaying exponentials and use the single stellar population of \cite{bruzual2003} with IMF~(Initial Mass Function) of \cite{chabrier2003}. We also take into account dust attenuation~\citep{calzetti2000} and dust emission~\citep{dale2014}. 
Although there are multi-wavelength data for the COSMOS and SXDS fields, such data are not available for some of our candidates (in particular those in the Deep layer).
Thus, we only use the HSC photometry to avoid possible systematics caused by the variety of the dataset.

\begin{figure}[ht]
  \begin{center}
    \includegraphics[scale=0.55]{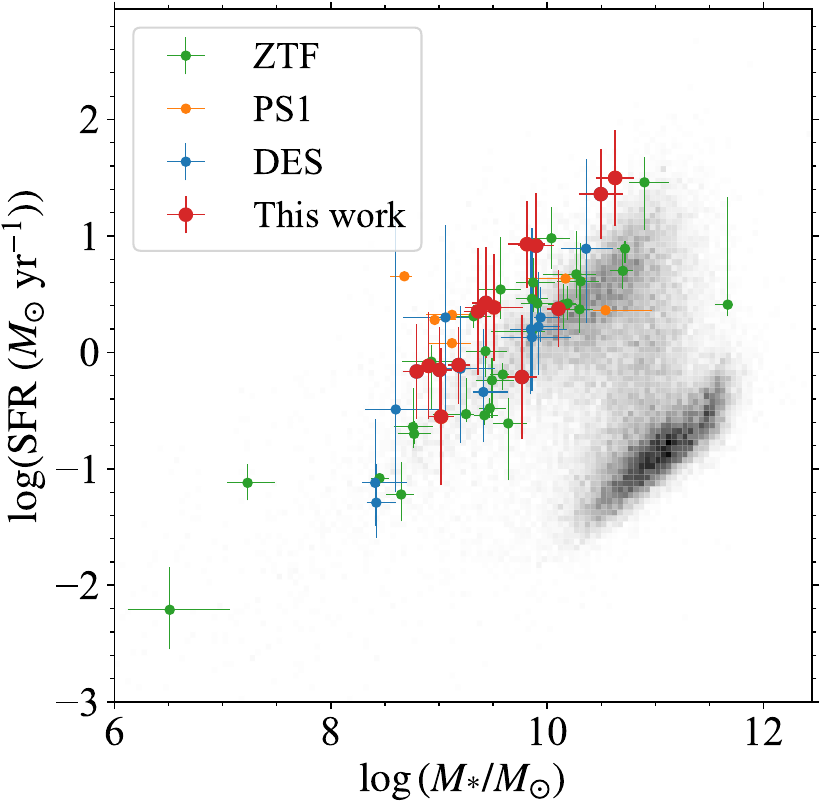}
\caption{The relation between stellar mass and SFR of host galaxies. Red points represent the rapid transient in our samples. We also show the host galaxy properties of rapid transient from PS1~\citep{drout2014}, DES~\citep{pursiainen2018} and ZTF~\citep{ho2023} with SDSS field galaxies~(gray points, DR7;~\citealt{abazajian2009}). 
 \label{fig:sfrvsM}
  }
\end{center}
\end{figure}

Figure~\ref{fig:sfrvsM} shows the relation between stellar mass and star formation rate~(SFR) of the host galaxies of our rapid transient samples~(red). We also show the properties of the host galaxy of rapid transient from PS1~\citep{drout2014}, DES~\citep{pursiainen2018}, and ZTF~\citep{ho2023} with field galaxies from the catalogues of SDSS Data Release 7~(DR7;~\citealt{abazajian2009}).

All the host galaxies of our samples show SFR higher than $0.1~M_{\odot} \ \rm{yr^{-1}}$ and lie on the star forming main-sequence of the SDSS field galaxies.
The stellar mass and SFR are also broadly consistent with those of the host galaxies of rapid transients ~\citep{wiseman2020}, supporting that our rapid transients link to massive stars.

The offsets of the transients from the centers of their host galaxies range from 0.4 to 17.1 \rm{kpc}~(Table~\ref{tab:hostgal}). It is interesting to note that three objects~(HSC17btum, HSC17dadp, and HSC17cgdi) have relatively large offsets~($\ge 10~\rm{kpc}$). These offsets are, however, still within the range of the observed offsets for core-collapse SNe~\citep{wang1997}. Note that a recently reported LFBOT, AT2023fhn~(the Finch), also has a large offset~(16.51~kpc) from its host galaxy~\citep{chrimes2024}.

\begin{table}[t]
  \caption{host galaxy properties}
  \label{tab:hostgal}
  \centering
  \begin{tabular}{lrrrrrclrc}
   	\hline 
	Name  &$\log{M_{*}}$ & $\log{\rm{SFR}}$ & 
 \multicolumn{2}{c}{Offset$^a$}\\
	 &$(M_{\odot})$
  & $(\rm{yr^{-1}})$ & (arcsec) & (kpc)& \\	
	\hline \hline
	HSC16apsu & 9.36 & 0.35 
 & 0.60 & $4.01\pm{0.13}$\\
	HSC17auls  & 9.18  & $-$0.11  
 & 0.81 &$3.75\pm{0.09}$\\ 
	HSC17bhyl  & 9.43 & 0.42 
 & 0.07 & $0.46\pm{0.13}$ \\ 
	HSC17btum & 10.10 &  0.37 
 & 2.14 & $13.13\pm{0.11}$\\ 
	HSC17dadp & 8.79 & $-$0.16 
 & 2.19 & $17.10\pm{0.14}$\\ 
	HSC17dbpf  &10.50  & 1.36 
 & 0.43 & $3.64\pm{0.16}$\\
	HSC17dhvg & 9.01 &  $-$0.15  
 & 0.08 & $0.48\pm{0.13}$\\ 
	HSC17dodm&  9.90 & 0.91  
 & 0.21 & $1.47\pm{0.16}$\\ 
      	\hline
	HSC17cgdi  & 9.77 & $-$0.21 
 & 1.67 & $12.99\pm{0.14}$\\ 	
	\hline
	HSC19aqfi  & 8.90 & $-$0.12 
 & 0.18 &$1.45\pm{0.14}$\\
	HSC19bfbh & 10.63  &1.50 
 & 0.58 &$4.85\pm{0.16}$\\ 
	HSC19bijv   & 9.51  & 0.39 
 & 0.13 &$1.15\pm{0.16}$\\ 
	\hline
	HSC19bbhh & 9.81  & 0.92 
 & 0.72 &$6.14\pm{0.16}$\\ 
	HSC19beav & 9.01  & $-$0.55  
 & 1.44 &$7.96\pm{0.09}$\\ 
	\hline
  \end{tabular}
  \tablecomments{
   $^a$ The error is propagation of typical uncertanity in offset $\sim 0.02$~\rm{arcsec}~\citep{aihara2018a} for each distance.}
\end{table}

\subsection{Classification based on color evolution and luminosity}

Here we try to further classify our rapid transients based on the color evolution and luminosity. As discussed in \citet{ho2023}, ``rapid transients'' consist of different types of the explosions, including early phase of Type IIb SNe, Type Ibn SNe, and LFBOTs.
As shown in the luminosity-duration diagram (Figure \ref{fig:phase}), our samples are not similar to LFBOTs.

Type Ibn SNe tend to be bluer and more luminous as compared with the early phase of Type IIb SNe. We estimate the $g-r$ color from the light curves of our rapid transient samples with the corresponding band in the rest frame. The $g-r$ color of Type IIb and Type Ibn SNe are also estimated from the light curves taken with ZTF, which are obtained through Lasair~\citep{smith2019}. 

\begin{figure}[ht]
  \begin{center}
    \includegraphics[scale=0.55]{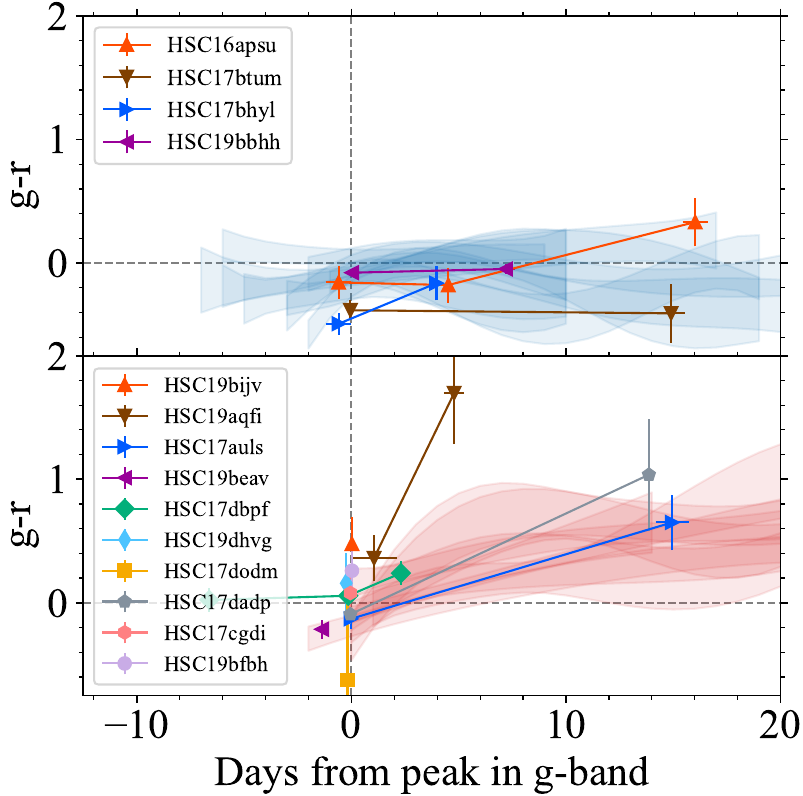}
\caption{$g-r$ color evolution of our rapid transient samples. The upper panel shows the our samples whose color evolve similarly to Type Ibn SNe, while the lower panel shows the our samples whose color evolve similarly to  normal SNe. We plotted the color evolution of Type Ibn SNe~(blue shaded region) and Type IIb SNe~(red shaded region) from ZTF~(Lasair;~\citealt{smith2019}) for comparisons in each panel.
  \label{fig:g-r}
  }
\end{center}
\end{figure}

The color evolution of our samples can be broadly divided into two types~(Figure \ref{fig:g-r}). Four objects~(HSC16apsu, HSC17btum, HSC17bhyl, and HSC19bbhh) show bluer color evolution until $\sim10$ days after the peak as in Type Ibn SNe. On the other hand, other four objects~(HSC17auls, HSC17dadp, HSC17dbpf, HSC19aqfi) show a redder color evolution similar to normal core-collapse SNe. Note that some objects show a blue color at the peak, we do not intend to classify them due to the limited light curve coverage. Among the objects with the blue color evolution, 
we approximately classify HSC16apsu and HSC17bhyl as Type Ibn SN candidates as they also show luminous peak magnitudes~($-18 \geq M_{\rm{peak}} \geq-20$) consistent with the typical light curve behavior of Type Ibn SNe~\citep{hosseinzadeh2018, ho2023}.

It is interesting to note that two of our samples~(HSC17dbpf and HSC19bbhh) show high luminosities ($M_{\rm{peak}}<-20$). 
Their properties are more similar to unusual Type Ic-BL SNe such as SN2018gep~\citep{ho2019} and iPTF16asu~\citep{whitesides2017, wang2019} or rapidly rising luminous transients~($t_{\rm{rise}}\sim 10$~day and $M_{\rm{peak}}\sim-20$~mag, \citealt{arcavi2016}) rather than LFBOTs like AT2018cow. 
In the luminosity and duration parameter space, these two objects are located in the short timescale end of SLSNe (Figure \ref{fig:phase}).

\subsection{Event rate}

In this subsection, we estimate the event rate of rapid transients at medium to high redshift range~($0.34 \leq z \leq 1.85$). In our analysis, we include both layers~(UD and D) for each field~(COSMOS and SXDS). We first estimate the approximated event rate  for each layer by the so called $1/V_{\rm{max}}$ method~\citep{schmidt1968}. Using redshift~$z$
and survey duration~$T$, the event rate $r$ of each object can be estimated as the function of absolute magnitude $M$:
\begin{align}
r(M) = \frac{1+z}{\epsilon_{\rm{det}}\epsilon_{\rm{sel}} TV_{\rm{max}}(M)}.
\end{align}
Here, $V_{\rm{max}}(M)$ is the comoving volume estimated with the maximum redshift at which an object with an absolute magnitude $M$ can be detected with our data. We take into account the survey efficiency by performing mock observations using \texttt{SNCosmo} as done in Section 3.1. Survey efficiency is expressed as a product of the efficiency of detection $\epsilon_{\rm{det}}$ and the efficiency of selection process $\epsilon_{\rm{sel}}$.
Since the luminosity function of rapid transients is poorly known, we perform mock observations by dividing our samples into three luminosity ranges (1)~$-17\geq M_{\rm peak}\geq-18$ mag, (2)~$-18\geq M_{\rm peak}\geq-19.5$ mag, and  (3)~$-20.5\geq M_{\rm peak}\geq-22$ mag. We assume a flat distribution in each range.
For the timescale, we limit the range to be $10~\rm{days}~$$\leq \Delta t_{1/2}\leq 18$ days so that the timescales of the template light curves match with those of the observed samples.
Then, we perform mock observations according to the observational schedule and depth of each field and layer (COSMOS UD/D, SXDS UD/D) and measure the fraction of transients satisfying the detection criteria of the survey ($5 \sigma$ detection more than twice), which gives a detection efficiency $\epsilon_{\rm{det}}$. Then, we also apply data quality cut (having data points before and after the peak) to the mock data, which is the major factor to affect the selection efficiency. A typical efficiency is estimated to be $\epsilon_{\rm{sel}} = 0.86~\rm{to}~1.00$. As a result, the survey efficiency for our samples in each field and layer is estimated to range from 0.29 to 0.86 depending on the field/layer.

After estimating the event rate in each field and layer, we derive the average event rate in each absolute magnitude bin. 
Note that we assume the brightest data point as a peak magnitude as in Figure \ref{fig:phase}. Since our light curves may have missed the true peak magnitude, the estimated event rate is subject to this uncertainty.

\begin{figure}[t]
  \begin{center}
    \includegraphics[scale=0.5]{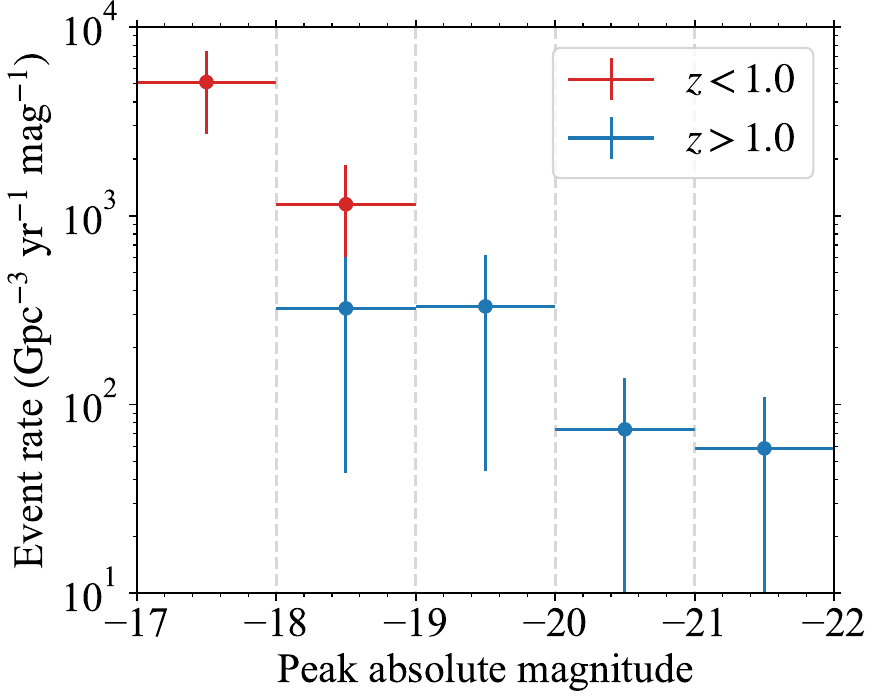}
\caption{Event rate as a function of the peak magnitude for the rapid transient from the Subaru HSC-SSP Transient Survey. We divided our samples into middle redshift samples~($z<1.0$, red) and high redshift samples~($z>1.0$, blue).
  \label{fig:rate}
  }
\end{center}
\end{figure}

Figure~\ref{fig:rate} shows the event rate as a function of the peak absolute magnitude. Considering the redshift evolution of cosmic star formation rate, we divide our samples into medium redshift samples~($z<1.0$) and high redshift samples~($z>1.0$). Although the conventional definition of ``rapid transient''  includes various types of transients~\citep{ho2023}, we estimate the event rate of the entire ``rapid transient'' from our survey to compare with those in the literature. For our medium redshift samples ($z<1.0$), the event rate of ``rapid transient'' is estimated to be $\sim6\times10^3$~$\rm{events~yr^{-1}~Gpc^{-3}}$ at a median redshift $z \sim 0.74$.
This is broadly consistent with the previous works: $\geq10^{3}~\rm{events~yr^{-1}~Gpc^{-3}}$ at a median redshift  $z\sim0.5$ by DES survey~\citep{pursiainen2018} and $4800$-$8000~\rm{events~yr^{-1}~Gpc^{-3}}$ at $z\sim0.3$ by PS1 survey~\citep{drout2014}.

 The event rate corresponds to $\sim 2$~\% of core-collapse SNe at $z=0.7$~($3.86^{+0.96}_{-0.72}\times 10^{5}~\rm{events~yr^{-1}~Gpc^{-3}}$;~\citealt{strolger2015}). This fraction is also consistent with the fraction estimated in \cite{pursiainen2018}. The fractions from PS1 and ZTF at low redshift are somewhat larger~(4-7~\% at $z\sim0.3$;~\citealt{drout2014} and 7~\% at $z\sim0.1$;~\citealt{ho2023}). 
 Since the event rate is dominated by low luminosity objects, the difference could be due to the magnitude limit~(Figure~\ref{fig:ZvsM}).

We also discuss the event rate of Type Ibn SN candidates. Although we cannot spectroscopically classify our samples into spectral types, we conservatively selected two Type Ibn SN candidates, HSC16apsu~($z=0.74$) and HSC17bhyl~($z=0.75$), based on the peak luminosities and $g-r$ color evolution (Section 5.2). We estimate the event rate for Type Ibn SNe candidates as $\sim6\times10^2~\rm{events~yr^{-1}~Gpc^{-3}}$. 

Type Ibn SNe are expected to originate from hydrogen-stripped massive stars with He-rich CSM that is eruptively ejected just before the explosion. Compared with the event rate of Type Ib SNe, which are the explosion of the hydrogen-stripped massive stars, we can estimate how often these eruptive mass loss occur in hydrogen-striped massive stars. In a low redshift range~($z\leq0.05$), the relative fraction of Type Ibn SNe to normal CCSNe can be estimated as $\sim 0.5$ \%~\citep{ho2023}. By using the relative fraction of Type Ib SNe to total CCSNe at $z\leq0.05$ ($\sim7.1$ \%;~\citealt{smith2011}, $\sim10.8$\%;~\citealt{shivers2017}), the relative fraction of Type Ibn SNe to Type Ib SNe is $\sim$5-7 \% in the local Universe. In a medium redshift range, the event rate of CCSNe is $3.86^{+0.96}_{-0.72}\times 10^{5}~\rm{events~yr^{-1}~Gpc^{-3}}$~\citep{strolger2015} at $z \sim 0.7$, which is similar to the redshift of our Type Ibn SN candidates~($z\sim0.75$). By assuming that the relative fraction of the Type Ib SNe to total CCSNe at this redshift range is the same as the fraction in the local Universe~(7-10 \%), the event rate of Type Ib SNe at $z\sim0.7$ is roughly estimated as (3-4) $\times 10^4~\rm{events~yr^{-1}~Gpc^{-3}}$. Compared with the event rate of Type Ibn SN candidates ($\sim 6\times10^2~\rm{events~yr^{-1}~Gpc^{-3}}$), the relative fraction of Type Ibn SNe to the Type Ib SNe is more than $\sim$1\%. 
This means that a fraction of hydrog5en-stripped envelope stars eruptively eject their helium envelope just before the explosion at medium redshift range~($z\sim0.7$) is roughly similar to that in the local Universe. 

Finally, we also estimate the event rate of rapidly evolving and luminous transients, which are located at short timescale end of SLSNe in the luminosity-duration diagram (Figure \ref{fig:phase}). From our two samples (HSC17dbpf at $z=1.851$ and HSC19bbhh at $z=1.370$), the event rate is estimated to be $\sim 1\times10^2~\rm{events~yr^{-1}~Gpc^{-3}}$. 
The event rates of SLSNe at high redshift have been estimated as $\sim400$~$\rm{events~yr^{-1}~Gpc^{-3}}$ at $z\sim2$~\citep{cooke2012} and $\sim900\pm520$~$\rm{events~yr^{-1}~Gpc^{-3}}$ at $z\sim2$~\citep{moriya2019}.
This suggests that a relatively large fraction (about $\sim10$-$25$ \%) of SLSNe form a subclass showing rapidly evolving light curves.

\section{conclusions}
\label{sec:conclusions}

We present a systematic search for rapid transients at medium to high redshifts in the Subaru HSC-SSP transient survey. By using on the photometric observational data of 3381 SN candidates in the COSMOS/SXDS field, we classified the objects into Type Ia, Ibc, II SNe, and rapid transients. After the quality cut and the visual inspection, we finally obtained 14 rapid transients including 4 objects already reported by~\cite{tampo2020}. 

Our rapid transient samples have a wide range of the redshift~($0.34 \leq z \leq 1.85$) and  peak absolute magnitude~($-17 \geq M_{\rm{peak}}\geq -22$). Their observational properties, as well as physical properties such as the photospheric radii and temperatures, are generally similar to the previously reported rapid transients~\citep{drout2014, pursiainen2018, ho2023}. 
Their host galaxies are all star forming galaxies, supporting a massive star origin.

We estimate the event rate of the entire ``rapid transient'' as a function of the absolute magnitude. The total rate is about $6\times10^3 ~\rm{events~yr^{-1}~Gpc^{-3}}$ at $z \sim 0.7$ and the rate is dominated by lower-luminosity objects. Among the rapid transient samples, we photometrically select two candidates for Type Ibn SNe. The lower limit of the event rate is $6\times10^2~\rm{events~yr^{-1}~Gpc^{-3}}$ which is  $\sim$ 1 \% of Type Ib SNe at a similar redshift. 
Two of our rapid transient samples show a high luminosity similar to SLSNe. Their event rate is $\sim 1\times10^2~\rm{events~yr^{-1}~Gpc^{-3}}$ at $z \sim 1.4$-$1.9$, which is about 10-25 \% of the event rate of SLSNe at $z \sim 2$.

\acknowledgments
This work was supported by JST FOREST Program (Grant Number JPMJFR212Y), the Grant-in-Aid for Scientific research from JSPS (20H00174, 20H00179, 22K14069, 23H00127, 23H04891, 23H04894, 24H00027, 24H01810, 24H01824, 24K00682). S.T. acknowledges support from Graduate Program on Physics for the Universe (GP-PU) at Tohoku University.

The Hyper Suprime-Cam (HSC) collaboration includes the astronomical communities of Japan and Taiwan, and Princeton University.  The HSC instrumentation and software were developed by the National Astronomical Observatory of Japan (NAOJ), the Kavli Institute for the Physics and Mathematics of the Universe (Kavli IPMU), the University of Tokyo, the High Energy Accelerator Research Organization (KEK), the Academia Sinica Institute for Astronomy and Astrophysics in Taiwan (ASIAA), and Princeton University.  Funding was contributed by the FIRST program from the Japanese Cabinet Office, the Ministry of Education, Culture, Sports, Science and Technology (MEXT), the Japan Society for the Promotion of Science (JSPS), Japan Science and Technology Agency  (JST), the Toray Science  Foundation, NAOJ, Kavli IPMU, KEK, ASIAA, and Princeton University. This paper is based [in part] on data collected at the Subaru Telescope and retrieved from the HSC data archive system, which is operated by Subaru Telescope and Astronomy Data Center (ADC) at NAOJ. Data analysis was in part carried out with the cooperation of Center for Computational Astrophysics (CfCA) at NAOJ.  We are honored and grateful for the opportunity of observing the Universe from Maunakea, which has the cultural, historical and natural significance in Hawaii. This paper makes use of software developed for Vera C. Rubin Observatory. We thank the Rubin Observatory for making their code available as free software at http://pipelines.lsst.io/. The Pan-STARRS1 Surveys (PS1) and the PS1 public science archive have been made possible through contributions by the Institute for Astronomy, the University of Hawaii, the Pan-STARRS Project Office, the Max Planck Society and its participating institutes, the Max Planck Institute for Astronomy, Heidelberg, and the Max Planck Institute for Extraterrestrial Physics, Garching, The Johns Hopkins University, Durham University, the University of Edinburgh, the Queen’s University Belfast, the Harvard-Smithsonian Center for Astrophysics, the Las Cumbres Observatory Global Telescope Network Incorporated, the National Central University of Taiwan, the Space Telescope Science Institute, the National Aeronautics and Space Administration under grant No. NNX08AR22G issued through the Planetary Science Division of the NASA Science Mission Directorate, the National Science Foundation grant No. AST-1238877, the University of Maryland, Eotvos Lorand University (ELTE), the Los Alamos National Laboratory, and the Gordon and Betty Moore Foundation. 

\software{Scikit-laern~\citep{scikit-learn}, SNCosmo~\citep{sncosmo}, MIZUKI~\citep{tanakamasayuki2015, tanakamasayuki2018}, CIGALE~\citep{boquien2019}}

\bibliography{reference}


\end{document}